\documentclass[UTF-8,preprintnumbers,superscriptaddress,aps,reprint,showkeys]{revtex4}
\usepackage{amsmath,amssymb,graphicx,bm}
\usepackage{subcaption}

\begin{document}
	\title{Observational properties of black hole in quantum fluctuation modified gravity}

\author{Jia-Jun Yin}
	\affiliation{College of Physics Science and Technology, Hebei University, Baoding 071002, China}	

\author{Tong-Yu He}
\affiliation{College of Physics Science and Technology, Hebei University, Baoding 071002, China}

\author{Ming Liu}
\affiliation{College of Physics Science and Technology, Hebei University, Baoding 071002, China}

\author{Hui-Min Fan}
\affiliation{College of Physics Science and Technology, Hebei University, Baoding 071002, China}

\author{Bohai Chen}
\affiliation{School of Liberal Arts and Sciences, North China Institute of Aerospace Engineering, Langfang 065000, China}

\author{Zhan-Wen Han}
\affiliation{College of Physics Science and Technology, Hebei University, Baoding 071002, China}
\affiliation{Yunnan Observatories, Chinese Academy of Sciences, Kunming 650216, China}

\author{Rong-Jia Yang \footnote{Corresponding author}}
\email{yangrongjia@tsinghua.org.cn}
\affiliation{College of Physics Science and Technology, Hebei University, Baoding 071002, China}
\affiliation{Hebei Key Lab of Optic-Electronic Information and Materials, Hebei University, Baoding 071002, China}
\affiliation{National-Local Joint Engineering Laboratory of New Energy Photoelectric Devices, Hebei University, Baoding 071002, China}
\affiliation{Key Laboratory of High-pricision Computation and Application of Quantum Field Theory of Hebei Province, Hebei University, Baoding 071002, China}

\begin{abstract}
 This study investigates the properties of the thin accretion disk around a black hole in quantum fluctuation modified gravity (QFMGBH) using the Novikov-Thorne model. We restrict the parameter \( \alpha \) characterizing the quantum fluctuation of metric and compute the ISCO radius, examining its effect on the  energy flux, the radiation temperature, the luminosity spectrum, the energy efficiency, and the shadow size for various values of the parameter \( \omega \) characterizing the matter around the black hole. Our findings show that the ISCO radius increases with \( \alpha \) for \( \omega = 1/3, 0, -2/3 \), but decreases for \( \omega = -4/3 \). As \( \alpha \) increases, the energy flux and the temperature show distinct trends, with changes in the luminosity spectrum and the efficiency. The shadow size increases for \( \omega = 1/3, 0, -2/3 \), and decreases for \( \omega = -4/3 \). The accretion disk around the QFMGBH is smaller, hotter, and brighter than that surrounded the Schwarzschild BH in GR, suggesting that the observable differences can potentially distinguish quantum fluctuation modified gravity from standard GR.
\end{abstract}
\maketitle

\section{Introduction}
Various observations have confirmed that the expansion of the universe is accelerating. Modified gravity theories, such as $f(R)$ gravity, $f(R, T)$ theory, and Gauss-Bonnet gravity, are popular attempts to explain this phenomenon. If considering Heisenberg's nonperturbative quantization \citep{Dzhunushaliev:2013nea,Dzhunushaliev:2015mva,Dzhunushaliev:2015hoa}: the metric is decomposed into the sum of the classical part and the quantum fluctuation part, the corresponding quantum Einstein gravity engenders modified gravity model at the classical level, which can be called as quantum fluctuation modified gravity (QFMG). QFMG has been realized in some special models, see for example \cite{Yang:2015jla,Liu:2016qfx,Bernardo:2021ynf,Chen:2021oal,Lima:2023qdv}. QFMG has some interesting properties, see for example, it will result in a non-minimal coupling between geometry and matter, which gives an possible microscopic quantum description of the matter creation processes in $f(R, T)$ or $f(R, L_{\rm{m}})$ gravity \cite{Liu:2016qfx,Yang:2020lxv}. Baryogenesis in QFMG was recently discussed in \cite{Yang:2024gnf}. Black hole (BH) solutions in QFMG were considered in \cite{Yang:2020lxv,Hua:2024jzx}. Friedmann-Lemaitre-Robertson-Walker cosmology in QFMG was investigated in  \cite{Yang:2015jla,Liu:2016qfx,Bernardo:2021ynf,Haghani:2021iqe,Chen:2021oal}. 

The observations of the shadows of the supermassive BHs at the centers of the galaxy M87* and the Milky Way by the Event Horizon Telescope (EHT) Collaboration \cite{EventHorizonTelescope:2019dse,EventHorizonTelescope:2022wkp}, has given a boost to the study of BH physics. BHs are usually surrounded by thin accretion disk. Due to the strong gravitational field, the thin accretion disk will radiate high-energy fluxes \cite{Collodel:2021gxu,Liu:2024brf,Wu:2024sng,Das:2024tyh,Abbas:2023rzk,He:2022lrc,Gyulchev:2019tvk,Gyulchev:2020cvo}, which have significant effects
on BH’s shadow. The study of BH's shadow has been ongoing for a long time.
In 1979, Luminet had obtained simulated images of a BH with thin accretion disk, illustrating that the BH's shadow depends on the accretion flow \cite{Luminet:1979nyg}. The shadow for Kerr BH with Keplerian accretion disk had been discussed in detail in \cite{Beckwith:2004ae}. Using the general relativistic ray-tracing code, reference \cite{Vincent_2011} had investigated the shadow images of a geometrically thick and infinite thin accretion disk around the compact objects. By considering the shape and the size of BH's shadow depending on the spacetime metric, the observation of BH provides an important way to extract useful information about the strong gravitational field, and helps us to distinguish various gravitational theories and even to investigate the quantum gravity effects. Recently a large nunber of studies on BH shadow have been made in various BH spacetimes, see for example \cite{Vagnozzi:2022moj,Uniyal:2022vdu,Heydari-Fard:2022xhr,Zare:2024dtf,Ma:2022jsy,Meng:2024puu,Fu:2022yrs,Mirzaev:2022xpz,Zhang:2024jrw,Zhang:2024lsf,Gyulchev:2024iel, Li:2022eue, Zhu:2024vxw, Luo:2024nul, Rodriguez:2024ijx, Macedo:2024qky, Hu:2023pyd, Meng:2023uws, Johnson:2023skw, Tang:2023lmr, Fauzi:2024nta, Zheng:2024ftk, Yang:2024nin, Gao:2023mjb, Zhang:2023okw, Vincent:2022fwj, Ban:2024qsa, Sui:2023tje, Benavides-Gallego:2024jyw, Liu:2024lbi}.

BH are important research subjects in gravitational theories, serving as a crucial test for our understanding of general relativity (GR) and modified gravity. Because of the presence of the matter-geometry coupling in QFMG, it is expected differences between BH solutions in GR and QFMG. Such differences may become more prominent for strong gravitational field. So it is important to investigate the effects imposed by QFMG in the scale of compact objects, such as BHs. In \cite{Hua:2024jzx}, an interesting BH solution of QFMG has been obtained, which has two characteristic parameters besides mass: one parameter characterizes the quantum fluctuation of metric, the other characterizes the matter surrounded the BH. Here we will discuss the effects of these two parameters on the thin accretion disk and the shadow of BH.

The paper is organized as follows. In Section II, we will briefly review the line element and model the motion of massive and massless particles respectively. In Section III we model thin accretion disk in
the geometry of the QFMGBH and use GYOTO, a general relativistic ray-tracing code \cite{Hua:2024jzx} to simulate the image of the QFMGBH’s shadow. Finally in section IV we summarize the results and their implications. Throughout the article we use geometrized units setting $c = 1$ with the metric signature \(-, +, +, +\).

\section{The spacetime and the motion of particles}
In this section, we will briefly introduce the BH solution in the QFMG and give the geodesic equations of moving particles in the equatorial plane. We will also discuss the range of the parameter $\alpha$ via the physical properties of the BH.

\subsection{Black hole in quantum fluctuation modified gravity}

In \cite{Hua:2024jzx}, an interesting QFMGBH solution for the gravitational field equations in QFMG proposed in \cite{Yang:2015jla} was obtained, which reduces to different classes of BHs surrounded by Kiselev fluids, by taking some specific values of the parameter of the equation of state (EoS). The geometry of the static and spherically symmetric QFMGBH in Schwarzschild coordinates is given by
\begin{equation}
\begin{split}
ds^2 &=-g_{tt}dt^2+g_{rr}dr^2+g_{\theta\theta}d\theta^2+g_{\phi\phi}d\phi^2 \\
&=-Bdt^2+\frac{1}{B}dr^2+r^2d\theta^2+r^2 \sin^2\theta d\phi^2 ,
\end{split}
\end{equation}
with
\begin{equation}
  \begin{split}
  B&=1-\frac{2M}{r}+Dr^{-\frac{2(1+3\omega-4\omega\alpha)}{2-3\alpha+\omega\alpha}},\\
 \end{split}
 \label{1r}
 \end{equation}
 where $D$ is an integration constant, the parameter $\alpha$ characterizing the fluctuation of metric, and the parameter $\omega$ denotes the EoS of  Kiselev fluid. If there are no fluctuations and fluids, solution \eqref{1r} will reduce to the Schwarzschild BH. Here, we mainly focus on the impacts of parameters $\omega$ and $\alpha$ on the research objects. For the convenience of calculations, we take $D=\lambda M$. The dimension of $\lambda$ depends on the parameters $\omega$ and $\alpha$, so dose the physical property of the parameter $D$, see for example, when $\omega=0$ and $\alpha=0$, $D$ is equivalent to mass. When the BH is surrounded by different fluids, such as radiation, dust, quintessence, cosmological constant, or phantom, solution \eqref{1r} reduces to different BH spacetime. 

\subsection{Asymptotically flat condition}

Now we discuss the range of the quantum fluctuations in the QFMGBH spacetime. We investigate five values of $\omega$ that are $\omega=1/3$, $\omega=0$, $\omega=-2/3$, $\omega=-1$, and $\omega=-4/3$, and these five values of $\omega$ represent the radiation field, the dust field, the quintessence field, the cosmological constant field, and the phantom field, respectively. 

\begin{figure}[htbp]
    \centering
    \begin{subfigure}[t]{0.4\textwidth}
        \centering
        \includegraphics[width=\textwidth]{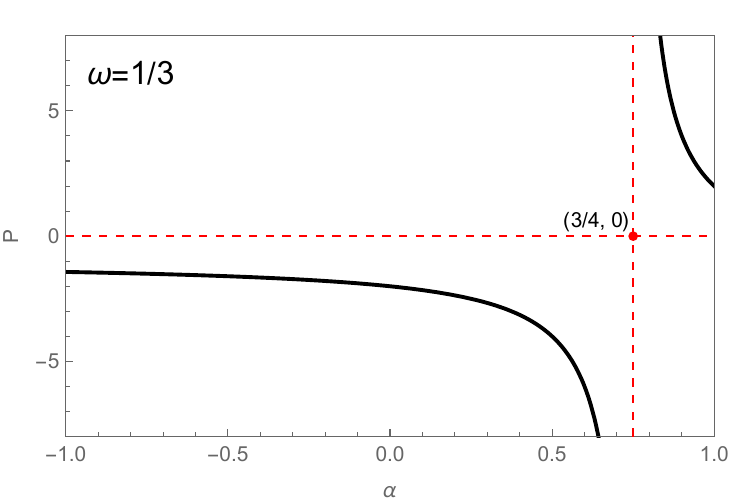} 
        \caption{ $\omega=1/3$}
        \label{1a}
    \end{subfigure}
    \hspace{0.75cm} 
    \begin{subfigure}[t]{0.4\textwidth}
        \centering
        \includegraphics[width=\textwidth]{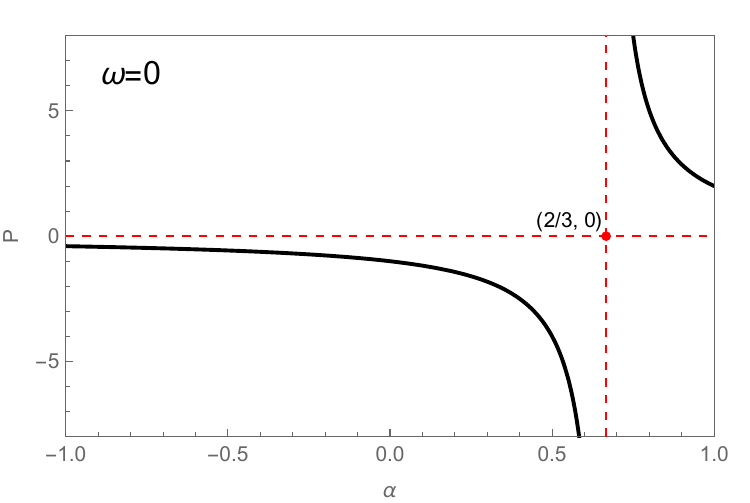} 
        \caption{ $\omega=0$}
        \label{1b}
    \end{subfigure}

    \vspace{0.2cm} 

    \begin{subfigure}[t]{0.4\textwidth}
        \centering
        \includegraphics[width=\textwidth]{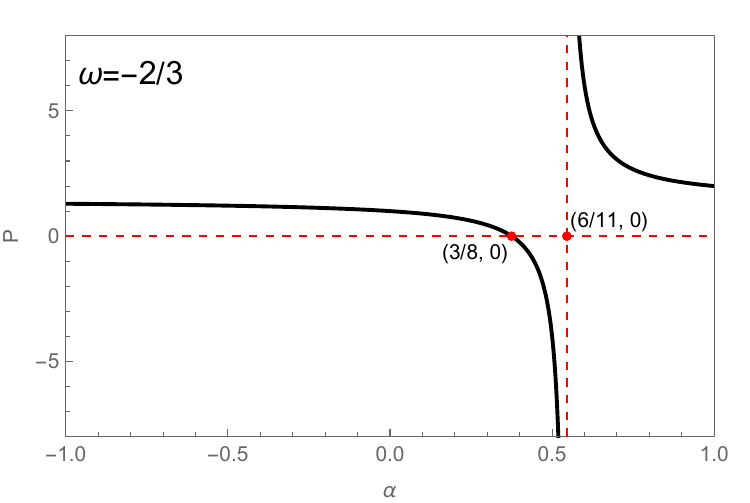} 
        \caption{ $\omega=-2/3$}
        \label{1c}
    \end{subfigure}
    \hspace{0.75cm} 
    \begin{subfigure}[t]{0.4\textwidth}
        \centering
        \includegraphics[width=\textwidth]{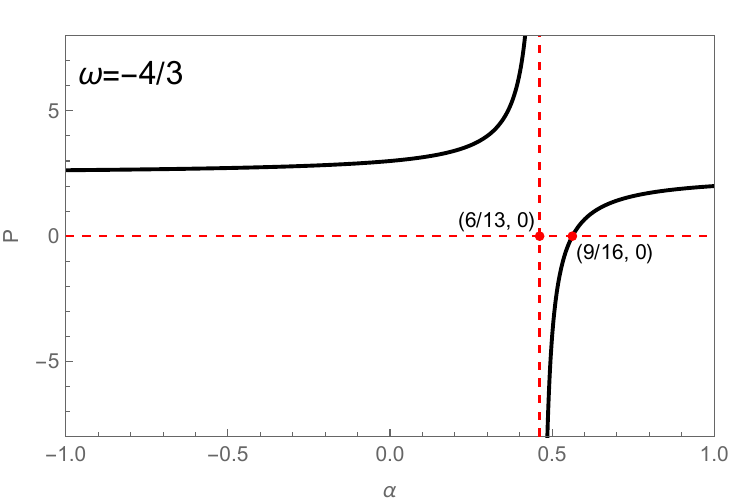} 
        \caption{ $\omega=-4/3$}
        \label{1d}
    \end{subfigure}

    \caption{ $P$ in the range where $\alpha$ belongs to (-1, 1) is plotted for different $\omega$. The intervals and coordinates where $P < 0$ are labeled.}
    \label{1}
\end{figure}

In the subsequent content, we will consider the Novikov-Thorne model for accretion, which bases on asymptotically flat spacetime. So we assume that the QFMGBH spacetime should be asymptotically flat at infinity, namely, it has to satisfy the condition $B\to 1$ when $r\to\infty$. In \cite{Yang:2015jla}, the range of the parameter $\alpha$ is $(-1,1)$. We can further constrain $\alpha$ by the asymptotically flat condition, The relevant inequality is shown as 
\begin{equation}
    P=-\frac{2(1+3\omega-4\omega\alpha)}{2-3\alpha+\omega\alpha}<0
    \label{2.10}.
\end{equation}
Through analyzing the inequality \eqref{2.10}, we can know that as $\alpha$ varies in range $(-1,1)$, the $P$ identically equals to $1$ for $\omega=-1$, which doesn't satisfy the asymptotically flat condition. So we don't consider this case ($\omega=-1$) further. In the same way, we can acquire the ranges of $\alpha$ for other different values of $\omega$: the range of $\alpha$ for $\omega=1/3$ is $(-1,3/4)$, for $\omega=0$ is $(-1,2/3)$, for $\omega=-2/3$ is $(3/8,6/11)$, and for $\omega=-4/3$ is $(6/13,9/16)$. We visualize these ranges in figure \ref{1}.

\subsection{Stable circular orbits}
From the definition of the BH, we can know that the BH cannot be directly observed, but we can understand the properties of the central objects by researching the properties of its surroundings \cite{Urmanov_2024}, so the time-like geodesics around compact astrophysical objects are very crucial. As usually, we assume that the accretion disk is located in the equatorial plane of the BH ($\theta=\pi/2$). The Lagrangian of particles moving in the equatorial plane of a static spherically symmetric BH is as follows
\begin{equation}
    \begin{split}
    \mathcal{L}&=\frac{1}{2}g_{\mu\nu}\dot{x^{\mu}}\dot{x^{\nu}}\\
    &=\frac{1}{2}g_{tt}\dot{t}^2+\frac{1}{2}g_{rr}\dot{r}^2+\frac{1}{2}g_{\phi\phi}\dot{\phi}^2,
    \end{split}
\end{equation}
where the dot means to take the derivative with respect to affine parameter $\lambda$. According to the Euler-Lagrangian equation, we can obtain 
\begin{equation}
    \begin{split}
        g_{tt}\dot{t}&=-E,\\
        g_{\phi\phi}\dot{\phi}&=L,
    \end{split}
\end{equation}
where $E$ and $L$ represent the specific energy and the specific angular momentum of particles. From the above two equations, we can obtain the components of the particle four-velocity that are $\dot{t}$ and $\dot{\phi}$, respectively
\begin{equation}
    \begin{split}
        \dot{t}&=-\frac{E}{g_{tt}},\\
        \dot{\phi}&=\frac{L}{g_{\phi\phi}}.
    \end{split}
\end{equation}
We can acquire the radial equation of motion via the normalization $g_{\mu\nu}\dot{x}^\mu\dot{x}^\nu=-1$,which is given by
\begin{equation}
    V_{\rm{eff}}(E,L,r)=-1+\frac{g_{\phi\phi}E^2+g_{tt}L^2}{-g_{tt}g_{\phi\phi}},
\end{equation}
where $V_{\rm{eff}}=g_{rr}\dot{r}^2$ denotes the effective potential of particles. Circular orbits exist at positions where $\dot{r} = 0$, which corresponds to the conditions $V_{\rm{eff}} = 0$ and $V_{\rm{eff}}^{\prime} = 0$ \cite{Gair_2008,Bambi_2011(1),Bambi_2011(2)}, where the prime denotes differentiation with respect to $r$. Under these conditions, we can derive expressions for the specific energy $E$, the specific angular momentum $L$, and the specific angular velocity $\Omega$ of particles in the equatorial plane of a Schwarzschild-like black hole spacetime.

\begin{equation}
    \Omega=\frac{d\phi}{dt}=\frac{\sqrt{-g_{tt,r}g_{\phi\phi,r}}}{g_{\phi\phi,r}},
    \label{2.7}
\end{equation}
\begin{equation}
     E=-\frac{g_{tt}}{\sqrt{-g_{tt}-g_{\phi\phi}\Omega^2}},
       \label{2.8}
\end{equation}
\begin{equation}
     L=\frac{g_{\phi\phi}\Omega}{\sqrt{-g_{tt}-g_{\phi\phi}\Omega^2}}.
       \label{2.9}
\end{equation}
We can derive the innermost stable circular orbit (ISCO) for particles in the circular orbits by putting the Eqs. \eqref{2.7}--\eqref{2.9} into $V_{\rm{eff}}^{\prime\prime}=0$. 

\section{Physical properties of thin accretion disks}
Next, we investigate the accretion process in thin disk around a QFMGBH. We will discuss in detail the effects of the parameters $\omega$ and $\alpha$ on the energy flux, the radiation temperature, the emission spectrum, the energy conversion efficiency, and the shadow (we use natural Units and take $\lambda=M=1$ here).   

Firstly, we provide a brief overview of the key properties of the accretion disk in the Novikov-Thorne model \cite{Shakura:1972te,Novikov:1973kta}. $(\textbf{1})$ The spacetime around the central mass object is static, axisymmetric, and asymptotically flat. $(\textbf{2})$ The gravitational effect that the accretion disk generate can be ignored. $(\textbf{3})$ The accretion disk is geometrically thin and optically thick, with the disk’s semi-thickness $h$ being significantly smaller than its radial extent $r$. As a result, light is absorbed when it passes vertically through the disk. $(\textbf{4})$ The inner edge of the disk is defined by the radius of the marginally stable orbit, with particles confined between the ISCO and the outer edge. $(\textbf{5})$ The accretion disk coincides with the equatorial plane of the compact accreting object. $(\textbf{6})$ The radiation emitted from the disk is modeled as a black-body spectrum. $(\textbf{7})$ The mass accretion rate $\dot{M}$ is constant over time.

\subsection{The radiant energy flux}
Considering the accretion disk around a QFMGBH, the metric components as well as the accretion particles move in Keplerian orbits with a rotational velocity $\Omega$, the specific energy $E$, and the specific angular momentum $L$ depends only on the radial coordinate. The particles with the four-velocity $u^{\mu}$ form a disk which has an averaged surface density $\Sigma$
\begin{equation}
   \Sigma=\int^{h}_{-h}\langle\rho_0\rangle dz,
\end{equation}
where $h$ is the disk height during the time $\Delta t$, $\rho_0$ is the density of rest mass, and $\langle A\rangle$ is defined as the averaged value of $A$ in time $\Delta t$ with angle $2\pi$. The particles in the disk is modeled by an anisotropic fluid source which reads
\begin{equation}
  T^{\mu\nu}=\rho_0u^{\mu}u^{\nu}+2u^{(\mu}q^{\nu)}+t^{\mu\nu},
\end{equation}
where the stress tensor $t^{\mu\nu}$ and the energy flow vector $q^{\mu}$ are measured in the averaged rest-frame with $u_{\mu}q^{\mu}=0$. Here the specific heat is neglected. The torque of the disk structure is given by
\begin{equation}
   W^{r}_{\phi}=\int^{h}_{-h}\langle t^{r}_{\phi}\rangle dz.
\end{equation}
The most important physical measure recounting the disk is the radiation flux $F(r)$ over the disk surface, which can be obtained from the time and orbital average of the energy flux vector
\begin{equation}
   F(r)=\langle q^{z}\rangle.
\end{equation}
Using stress energy tensor $T^{\mu\nu}$ of the disk, we can define the four-vectors of the energy and of the angular momentum flux in a local basis, respectively, as
\begin{equation}
E^\mu=-T^{\mu\nu}\left(\frac{\partial}{\partial t}\right)^{\nu}, ~~{\rm and}~~ J^\mu=-T^{\mu\nu}\left(\frac{\partial}{\partial\phi}\right)^{\nu}.
\end{equation}
The structure equations of the disk could be obtained from the conservation laws of the rest mass, of the energy, and of the angular momentum. The conservation for the rest mass, $\nabla_\mu(\rho_0u^{mu})=0$, yields the time averaged rate of the rest mass accretion, which is independent of the disk radius
\begin{equation}
  \dot{M}_0\equiv -2\pi\sqrt{-g}\Sigma u^{r}={\rm constant},
\end{equation}
where dot denotes the differentiation with respect to the time coordinate. The conservation for the energy, $\nabla_{\mu}u^{\mu}=0$, yields
\begin{equation}
  \left[\dot{M}_0E-2\pi\sqrt{-g}\Omega W^{r}_{\phi}\right]_{,r}=4\pi\sqrt{-g}FE,
\end{equation}
where the comma denotes the derivative with respect to the radial coordinate. This is an equilibrium equation, implying that the energy transported by the rest mass flow, $\dot{M}_0E$, and the energy transported by the torque in the disk,
$2\pi\sqrt{-g}\Omega W^{r}_{\phi}$, is well-adjusted by the energy emitted from the surface of the disk, $4\pi\sqrt{-g}FE$. The conservation for angular momentum, $\nabla_{\mu}J^{\mu}$, yields
\begin{equation}
  \left[\dot{M}_0L-2\pi\sqrt{-g} W^{r}_{\phi}\right]_{,r}=4\pi\sqrt{-g}FL,
\end{equation}
where the first term characterizes the angular momentum from the rest mass of the disk, the second term is the angular momentum transported by the torque in the disk, the right hand side term is the angular momentum
conveyed from the disk’s surface by radiation. Utilizing the above two equations, we can obtain the expression for the energy flux $F(r)$ by eliminating $W^{r}_{\phi}$. Using the relation between the energy and the angular momentum, $dE = QdJ$, for which 
the circular geodesic orbits assumes the form $E_{,r} = \Omega L_{,r}$, we can find the flux $F(r)$ of the radiant energy over the disk in terms of the specific energy, of the angular momentum and of the angular velocity of the disk matter in spherical coordinates
\begin{equation}
    F(r)=-\frac{\dot{M_{0} } \Omega_{,r}}{4\pi\sqrt{-g/g_{\theta\theta}}(E-\Omega L)^2}\int_{r_{\rm{ISCO}}}^r (E-\Omega L) L_{,r} \,dr,
\end{equation} 
where the tensor transformation from the original vertical flux $q^z$ (in cylindrical coordinates), $F(r)=\langle q^{z}\rangle=\langle q^{\theta}(r, \theta_0)\sqrt{g_{\theta\theta}(r, \theta_0)}\rangle$, when $z\ll r$ or $\theta_0\approx \pi/2$, has been used \cite{Collodel_2021}.

The QFMGBH cannot be directly observed, but the physical properties of the accretion disks around it are significant. We can further limit the range of $\alpha$ by the physical properties of the disks, for example the radiant energy flux. In other words, the numerical values of the radiant energy flux should be real numbers, not complex numbers. We find that the limited range of $\alpha$ for $\omega=1/3$ is $(-1.00,0.74]$, for $\omega=0$ is $[0,0.66]$, for $\omega=-2/3$ is $[0.45,0.54]$, and for $\omega=-4/3$ is $[0.47,0.53]$. These limited ranges also fulfill the asymptotically flat condition. Here we note that: $(\textbf{1})$ when $\omega=0$ and $\alpha=0$, the QFMGBH spacetime becomes the Schwarzschild-like spacetime with effective mass $2M_{\rm{eff}}=2M-D$; $(\textbf{2})$ when $\alpha=0.50$, as $\omega$ changes, $P$ always equals $-4$, and the QFMGBH spacetime remains unchanged. 

\begin{figure}[htbp]
    \centering
    \begin{subfigure}[t]{0.4\textwidth}
        \centering
        \includegraphics[width=\textwidth]{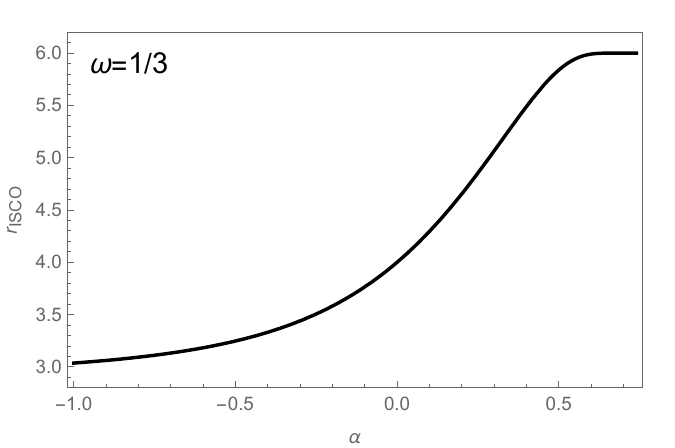} 
        \caption{ $\omega=1/3$}
        \label{2a}
    \end{subfigure}
    \hspace{0.75cm} 
    \begin{subfigure}[t]{0.4\textwidth}
        \centering
        \includegraphics[width=\textwidth]{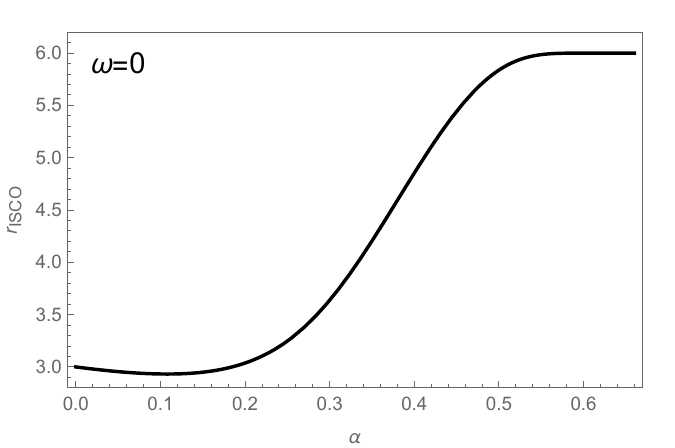} 
        \caption{ $\omega=0$}
        \label{2b}
    \end{subfigure}

    \vspace{0.2cm} 

    \begin{subfigure}[t]{0.4\textwidth}
        \centering
        \includegraphics[width=\textwidth]{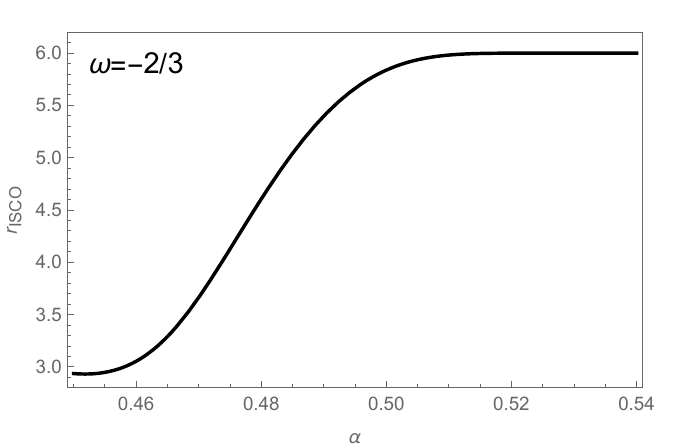} 
        \caption{ $\omega=-2/3$}
        \label{2c}
    \end{subfigure}
    \hspace{0.75cm} 
    \begin{subfigure}[t]{0.4\textwidth}
        \centering
        \includegraphics[width=\textwidth]{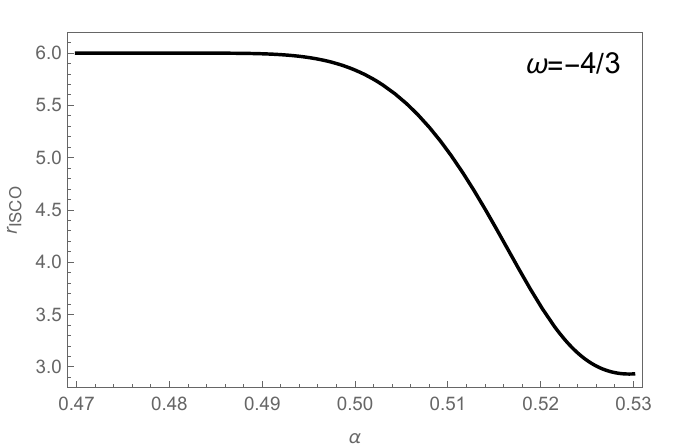} 
        \caption{ $\omega=-4/3$}
        \label{2d}
    \end{subfigure}

    \caption{The curves illustrate the variation of the innermost stable circular orbit radius, $r_{\rm{ISCO}}$, with respect to $\alpha$ for different values of $\omega$.
    }
    \label{2}
\end{figure}

Using the constrained ranges, the variation of $r_{\rm{ISCO}}$ is shown in figure \ref{2} from which we observe that $r_{\rm{ISCO}}\leq 6$, meaning that the innermost stable circular orbit radius of a thin disk around a QFMGBH is smaller than or equal to that of Schwarzschild BH. This is a significant result, which will impact the physical properties of the disks.

\begin{figure}[htbp]
    \centering
    \begin{subfigure}[t]{0.4\textwidth}
        \centering
        \includegraphics[width=\textwidth]{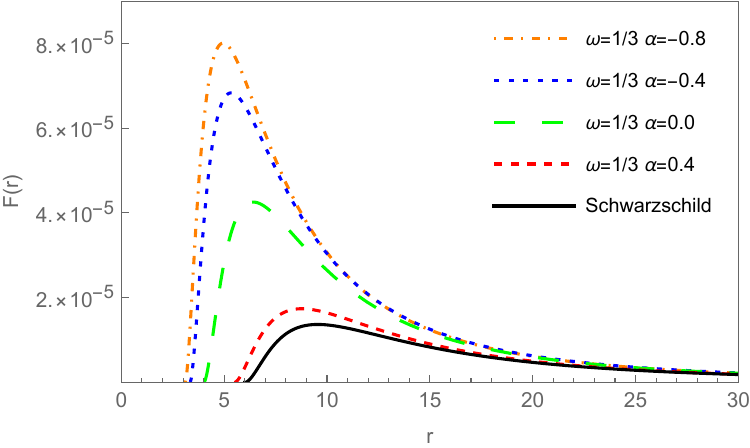} 
        \caption{ $\omega=1/3$}
        \label{3a}
    \end{subfigure}
    \hspace{0.75cm} 
    \begin{subfigure}[t]{0.4\textwidth}
        \centering
        \includegraphics[width=\textwidth]{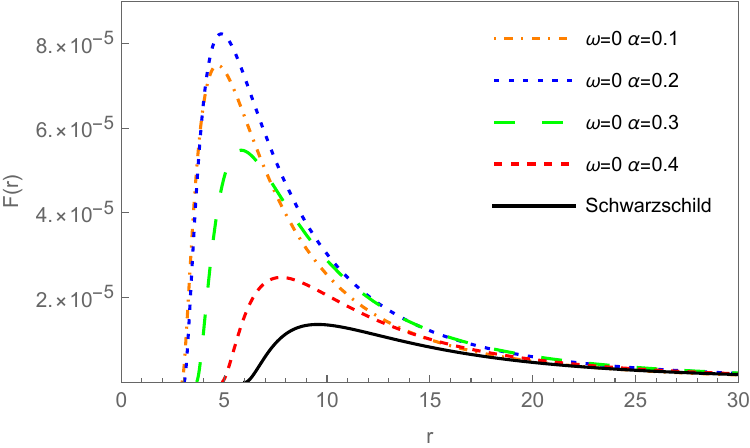} 
        \caption{ $\omega=0$}
        \label{3b}
    \end{subfigure}

    \vspace{0.2cm} 

    \begin{subfigure}[t]{0.4\textwidth}
        \centering
        \includegraphics[width=\textwidth]{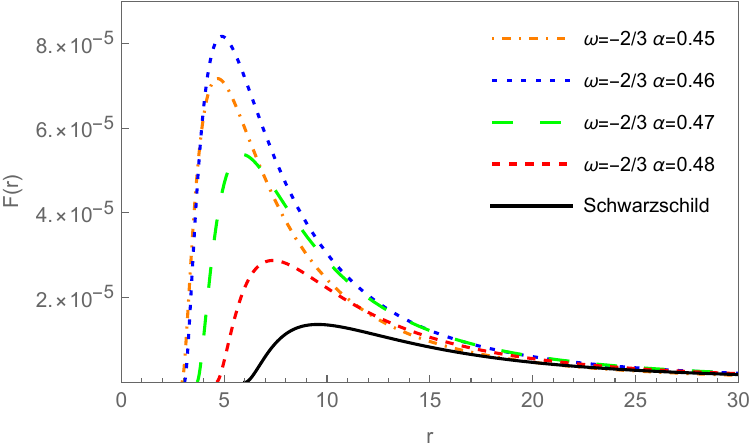} 
        \caption{ $\omega=-2/3$}
        \label{3c}
    \end{subfigure}
    \hspace{0.75cm} 
    \begin{subfigure}[t]{0.4\textwidth}
        \centering
        \includegraphics[width=\textwidth]{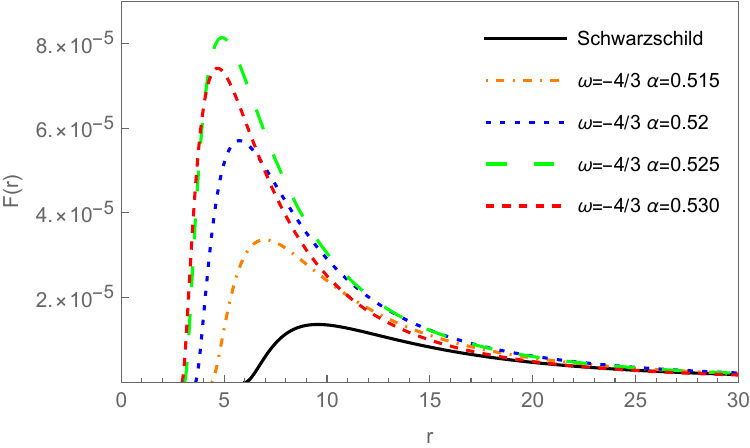} 
        \caption{ $\omega=-4/3$}
        \label{3d}
    \end{subfigure}

    \caption{ The radiant energy flux $F(r)$ for different values of $\omega$.}
    \label{3}
\end{figure}

\begin{figure}[htbp]
    \centering
    \begin{subfigure}[t]{0.4\textwidth}
        \centering
        \includegraphics[width=\textwidth]{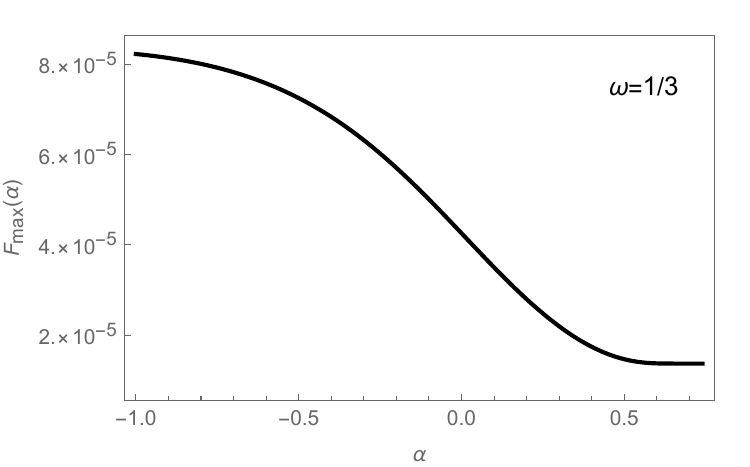} 
        \caption{ $\omega=1/3$}
        \label{4a}
    \end{subfigure}
    \hspace{0.75cm} 
    \begin{subfigure}[t]{0.4\textwidth}
        \centering
        \includegraphics[width=\textwidth]{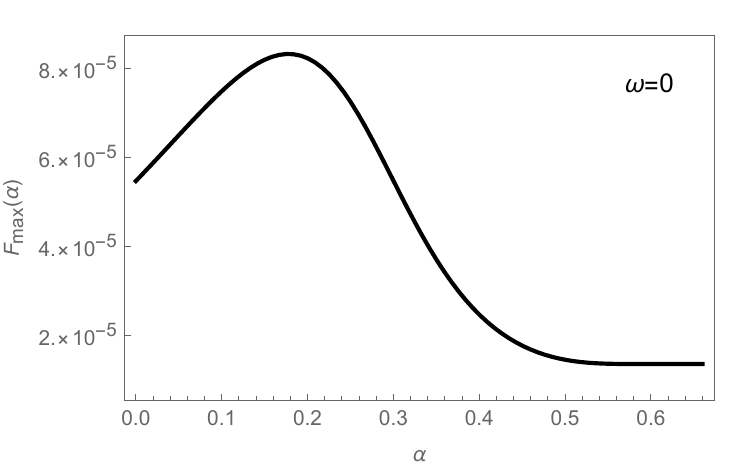} 
        \caption{ $\omega=0$}
        \label{4b}
    \end{subfigure}

    \vspace{0.2cm} 

    \begin{subfigure}[t]{0.4\textwidth}
        \centering
        \includegraphics[width=\textwidth]{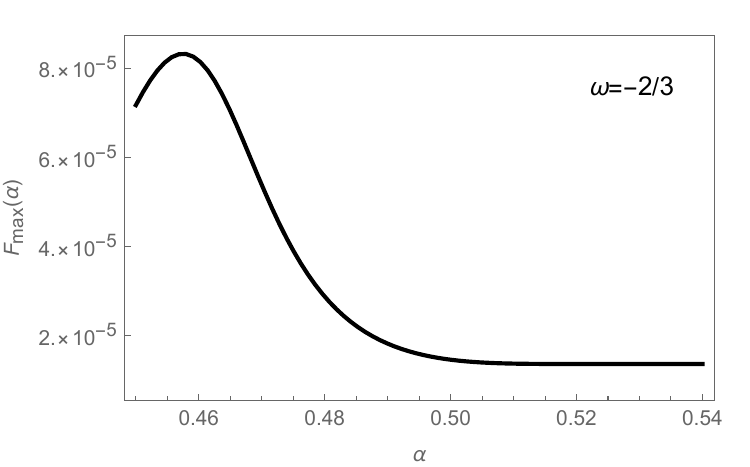} 
        \caption{ $\omega=-2/3$}
        \label{4c}
    \end{subfigure}
    \hspace{0.75cm} 
    \begin{subfigure}[t]{0.4\textwidth}
        \centering
        \includegraphics[width=\textwidth]{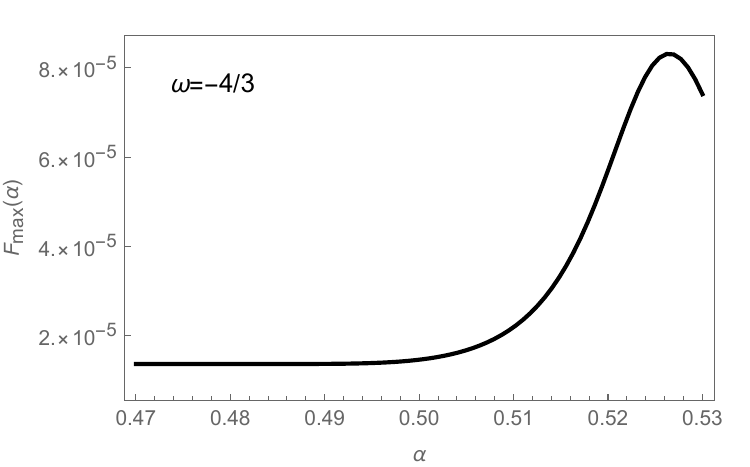} 
        \caption{ $\omega=-4/3$}
        \label{4d}
    \end{subfigure}
s
    \caption{The variation curves of the maximum energy radiant flux $F_{max}$ with respect to $\alpha$ for different values of $\omega$.}
    \label{4}
\end{figure}

The energy flux for different values of $\omega$ changes as $\alpha$ varies, as shown in figure \ref{3}. The maximum energy flux curves for different values of $\omega$ are presented in figure \ref{4}. From figures \ref{2}(a) and \ref{3}(a), we observe that as $\alpha$ increases, $r_{\rm{ISCO}}$ also increases, but the energy flux decreases simultaneously for $\omega = 1/3$. In figures \ref{2}(b) and \ref{3}(b), for $\omega = 0$, it is clear that as $\alpha$ increases, $r_{\rm{ISCO}}$ increases. By analyzing figures \ref{3} (b) and \ref{4}(b), we find that the maximum energy flux increases initially and then decreases as $\alpha$ increases. 

The behavior of the energy flux in figures\ref{2}(c) , \ref{3}(c) and \ref{4}(c) is similar to that in figures \ref{2}(b) ,\ref{3}(b) and \ref{4}(b), but the increase in energy flux for $\omega = -2/3$ occurs over a smaller range of $\alpha$ compared to that for $\omega = 0$. From figures \ref{2}(d) ,\ref{3}(d) and \ref{4}(d), for $\omega = -4/3$, we find that $r_{\rm{ISCO}}$ decreases as $\alpha$ increases, while the energy flux increases over a relatively large value and then decreases over a smaller range. Another interesting observation from figures \ref{3} and \ref{4} is that, for all four values of $\omega$, the energy fluxes exceed the energy flux of the Schwarzschild BH. Additionally, as $\alpha$ increases, for $\omega = 1/3$, $\omega = 0$, and $\omega = -2/3$, the radius of the maximum flux shifts outward; however, for $\omega = -4/3$, the radius of the maximum flux shifts inward. Moreover, the maximum energy flux for all four values of $\omega$ approaches a limit of $1.37 \times 10^{-5}$.

From these discussions, we see that there is degeneracy between parameters $\omega$ and $\alpha$. The general approach to reducing degeneracies between parameters is to fix some parameters while examining the impacts of changes in other parameters, as we have done in the analysis of the graph. In \cite{Chang:2024cbi}, Bayesian analysis was adopted to constrain the parameters based on the analytic dual-cone accretion model, however, some parameters were also fixed at the same time. In \cite{vanderGucht:2019bhl,Fang:2024hbw,Janssen:2025rbl,Sharipov:2025dgh}, machine learning methods were used to limit parameters, and generally some parameters were needed to be fixed in order to break the degeneracy between parameters, see for example \cite{vanderGucht:2019bhl}.

\subsection{The radiation temperature}
Then we investigate the radiation temperature $T(r)$ of the accretion disk. Since we adopt the steady-state thin disk model and assume this disk is in thermal equilibrium, we treat the radiation emitted by the accretion disk as perfect black-body radiation. The Stefan-Boltzmann law \( F(r) = \sigma_{\rm{SB}} T^4(r) \), where \( \sigma_{\rm{SB}} \) is the Stefan-Boltzmann constant, determines the radiation temperature \( T(r) \) of the accretion disk.

\begin{figure}[htbp]
    \centering
    \begin{subfigure}[t]{0.4\textwidth}
        \centering
        \includegraphics[width=\textwidth]{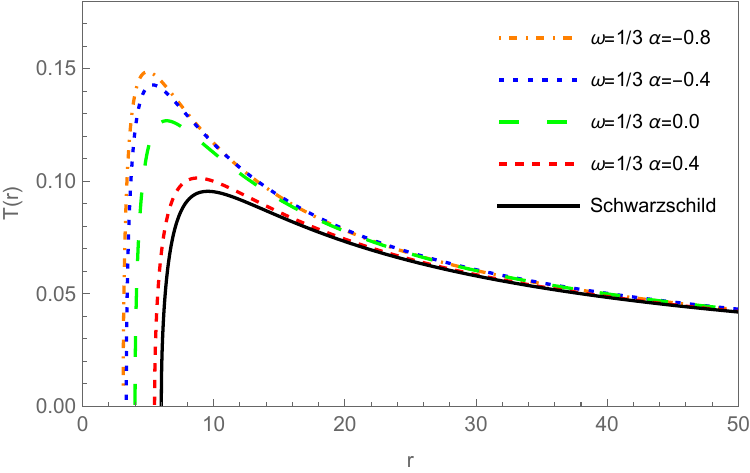} 
        \caption{ $\omega=1/3$}
        \label{5a}
    \end{subfigure}
    \hspace{0.75cm} 
    \begin{subfigure}[t]{0.4\textwidth}
        \centering
        \includegraphics[width=\textwidth]{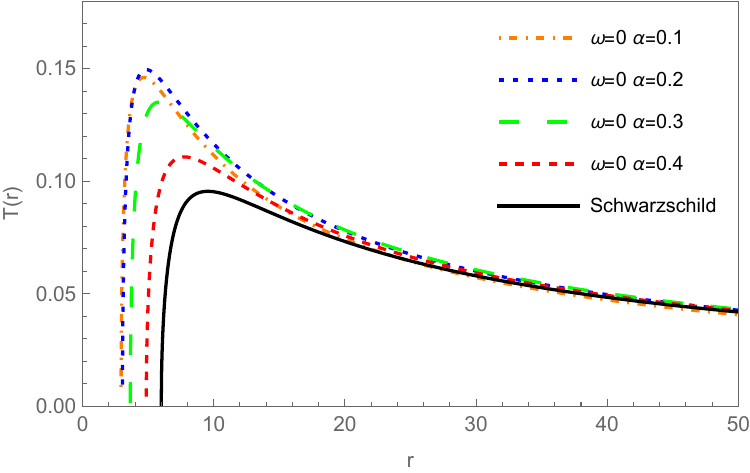} 
        \caption{ $\omega=0$}
        \label{5b}
    \end{subfigure}

    \vspace{0.2cm} 

    \begin{subfigure}[t]{0.4\textwidth}
        \centering
        \includegraphics[width=\textwidth]{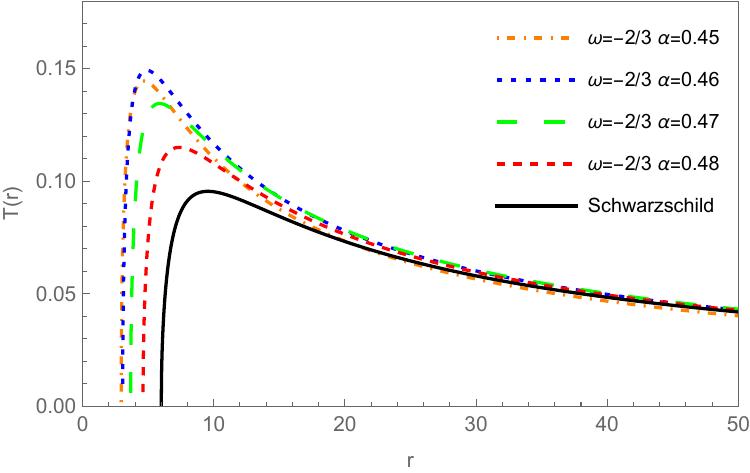} 
        \caption{ $\omega=-2/3$}
        \label{5c}
    \end{subfigure}
    \hspace{0.75cm} 
    \begin{subfigure}[t]{0.4\textwidth}
        \centering
        \includegraphics[width=\textwidth]{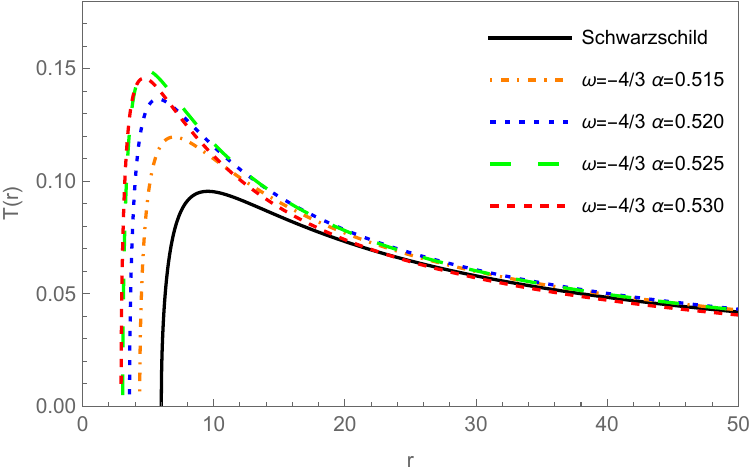} 
        \caption{ $\omega=-4/3$}
        \label{5d}
    \end{subfigure}

    \caption{The radiant temperature $T(r)$ for different values of  $\omega$.}
    \label{5}
\end{figure}

\begin{figure}[htbp]
    \centering
    \begin{subfigure}[t]{0.4\textwidth}
        \centering
        \includegraphics[width=\textwidth]{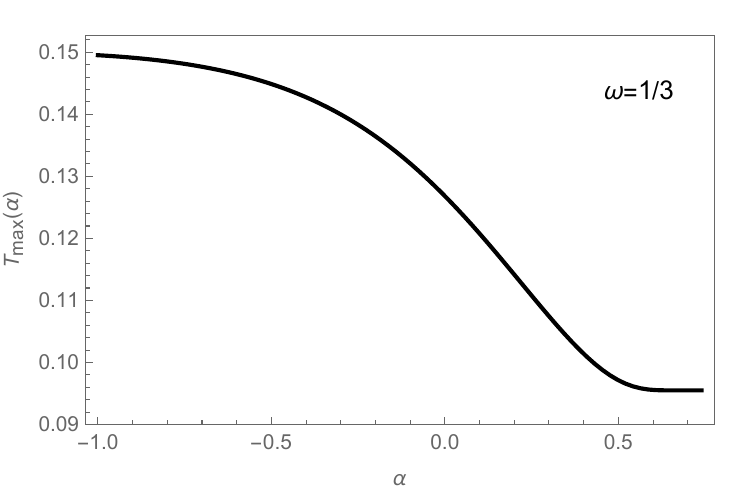} 
        \caption{ $\omega=1/3$}
        \label{6a}
    \end{subfigure}
    \hspace{0.75cm} 
    \begin{subfigure}[t]{0.4\textwidth}
        \centering
        \includegraphics[width=\textwidth]{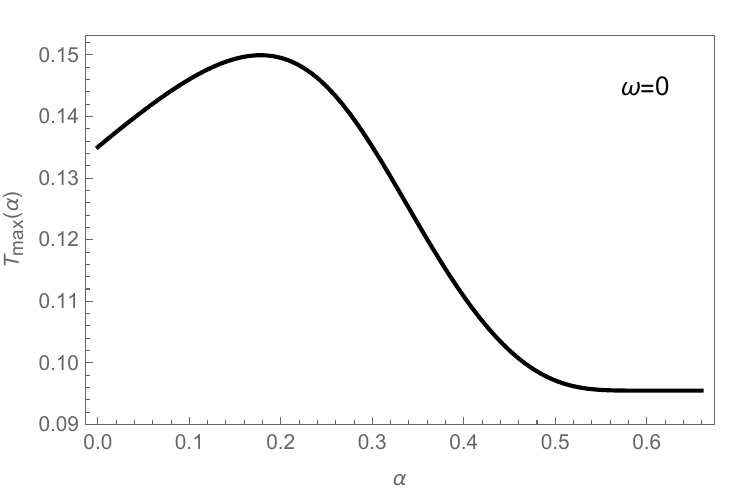} 
        \caption{ $\omega=0$}
        \label{6b}
    \end{subfigure}

    \vspace{0.2cm} 

    \begin{subfigure}[t]{0.4\textwidth}
        \centering
        \includegraphics[width=\textwidth]{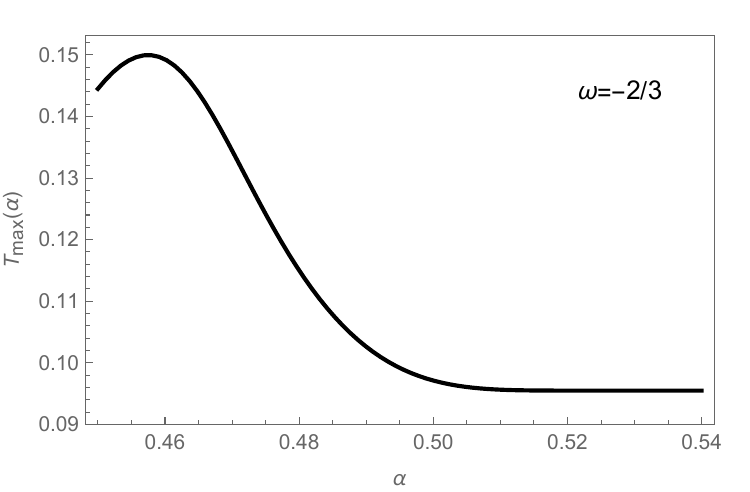} 
        \caption{ $\omega=-2/3$}
        \label{6c}
    \end{subfigure}
    \hspace{0.75cm} 
    \begin{subfigure}[t]{0.4\textwidth}
        \centering
        \includegraphics[width=\textwidth]{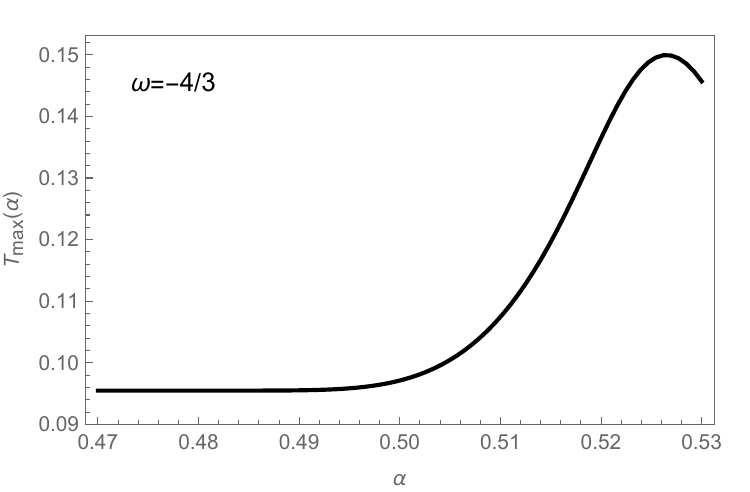} 
        \caption{ $\omega=-4/3$}
        \label{6d}
    \end{subfigure}

    \caption{The variation curves of the maximum radiant temperature \( T_{\text{max}}(\alpha) \) for different values of \( \omega \).}
    \label{6}
\end{figure}

Figures \ref{5} and \ref{6} show the radiation temperature on the surface of the accretion disk for different parameter values, as well as the change in the maximum radiation temperature for various values of \(\omega\) as \(\alpha\) increases. Analyzing figures \ref{5}(a) and \ref{6}(a) together, for \(\omega = 1/3\), we observe that as \(\alpha\) increases, the radiation temperature decreases. From figures \ref{5}(b) and \ref{6}(b), we find that for \(\omega = 0\), the radiation temperature initially increases with \(\alpha\), and decreases later. The trend in the radiation temperature resembles that of the energy flux. Figures \ref{5}(c) and \ref{6}(c) reveal that as \(\alpha\) increases, the change in radiation temperature for \(\omega = -2/3\) is similar to the \(\omega = 0 \). Figures \ref{5}(d) and \ref{6}(d) indicate that the variation in radiation temperature for \(\omega = -4/3\) is opposite to that of the energy flux for \(\omega = -2/3\). From figures \ref{5} and \ref{6}, we also observe that the change in the radius of the maximum values mirrors the behavior seen in the energy flux. Notably, the disk around the QFMGBH is hotter than that around the Schwarzschild BH. Furthermore, we find that the maximum radiation temperature for different values of \(\omega\) approaches a limit of \(0.095\).

\subsection{The observed luminosity} 

The luminosity \( L(\nu) \) of a thin accretion disk surrounding a BH is inferred from the red-shifted black-body spectrum and can be defined as \cite{Torres_2002}.
\begin{equation}
    L(\nu)=4\pi d^2 I(\nu)=\frac{8\pi h \cos\gamma}{c^2} \int_{r_{\rm i}}^{r_{\rm f}}\int_0^{2\pi}\frac{\nu_{\rm e}^3 r}{e^{\frac{h \nu_{\rm e}}{k_{\rm B} T}}-1}\,d\phi \,dr
\end{equation}
where \( d \) represents the distance from the disk's center, \( I(\nu) \) denotes the heat flux emitted by the accretion disk, \( h \) stands for the Planck constant, \( k_{\rm B} \) refers to the Boltzmann constant, and \( \gamma \) indicates the disk's inclination. We take \( h = 2\pi \) and \( k_{\rm B} = 1 \) in the natural unit system and \( \gamma = 0 \) in numerically calculations. The variables \( r_{\rm f} \) and \( r_{\rm i} \) in the integral represent the outer and inner radii of the accretion disk, respectively. We assume that the energy flux on the disk surface is zero at infinity; hence, we set \( r_{\rm f} \to \infty \) and \( r_{\rm i} = r_{\rm ISCO} \). By using the relation \( \nu_{\rm e} = \nu(1+z) \), we can obtain the emitted frequency. The redshift factor \( z \) is given as follows

\begin{equation}
    1+z=\frac{1+\Omega r \sin\phi \sin\gamma}{\sqrt{-g_{tt}-\Omega^2 g_{\phi \phi}}}
\end{equation}
The bending effect of light is ignored\cite{article,Bhattacharyya_2001}.

\begin{figure}[htbp]
    \centering
    \begin{subfigure}[t]{0.4\textwidth}
        \centering
        \includegraphics[width=\textwidth]{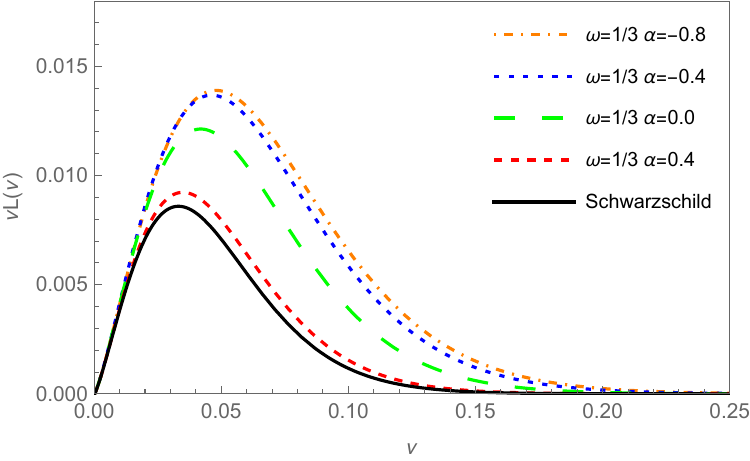} 
        \caption{ $\omega=1/3$}
        \label{7a}
    \end{subfigure}
    \hspace{0.75cm} 
    \begin{subfigure}[t]{0.4\textwidth}
        \centering
        \includegraphics[width=\textwidth]{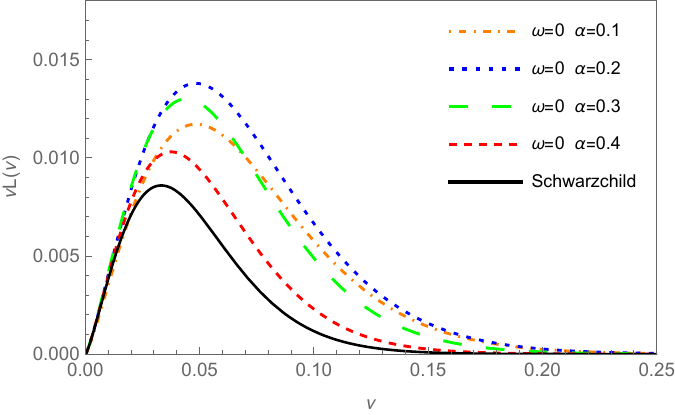} 
        \caption{ $\omega=0$}
        \label{7b}
    \end{subfigure}

    \vspace{0.2cm} 

    \begin{subfigure}[t]{0.4\textwidth}
        \centering
        \includegraphics[width=\textwidth]{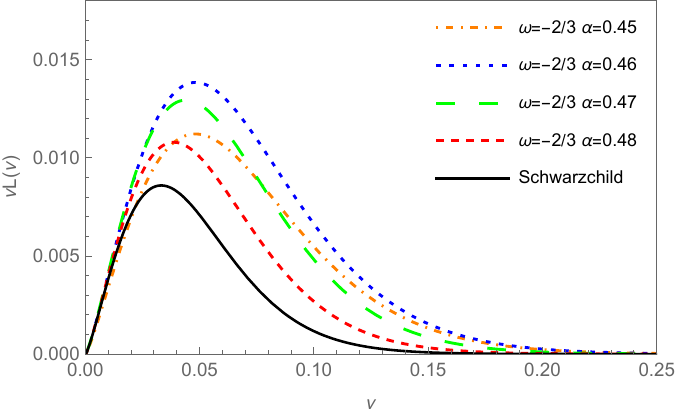} 
        \caption{ $\omega=-2/3$}
        \label{7c}
    \end{subfigure}
    \hspace{0.75cm} 
    \begin{subfigure}[t]{0.4\textwidth}
        \centering
        \includegraphics[width=\textwidth]{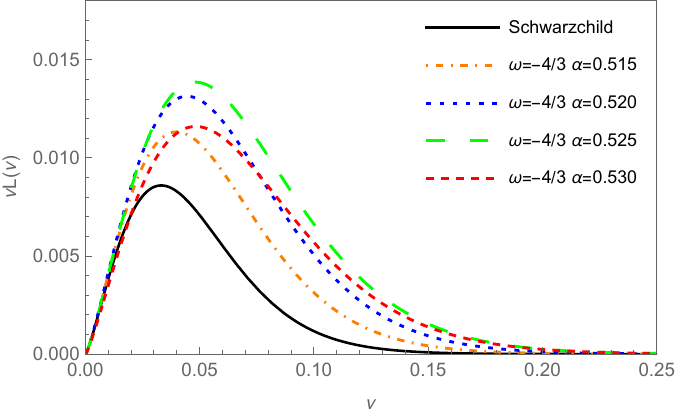} 
        \caption{ $\omega=-4/3$}
        \label{7d}
    \end{subfigure}
    \caption{The emission spectrum $\nu L(\nu)$ for different values of $\omega$.}
    \label{7}
\end{figure}

\begin{figure}[htbp]
    \centering
    \begin{subfigure}[t]{0.4\textwidth}
        \centering
        \includegraphics[width=\textwidth]{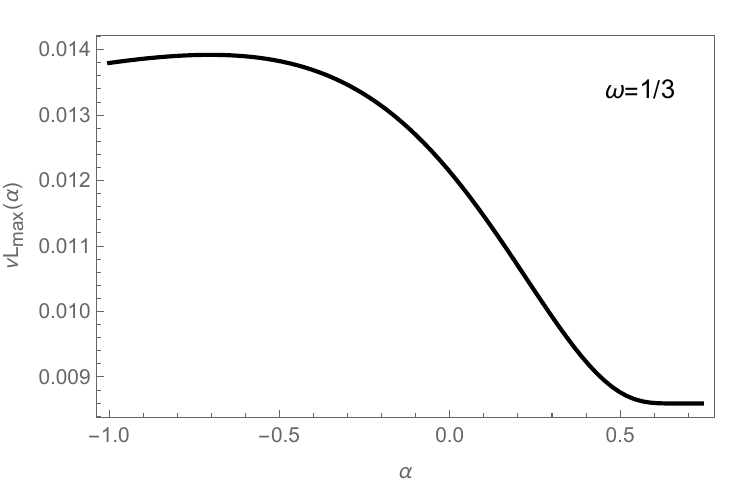} 
        \caption{ $\omega=1/3$}
        \label{8a}
    \end{subfigure}
    \hspace{0.75cm} 
    \begin{subfigure}[t]{0.4\textwidth}
        \centering
        \includegraphics[width=\textwidth]{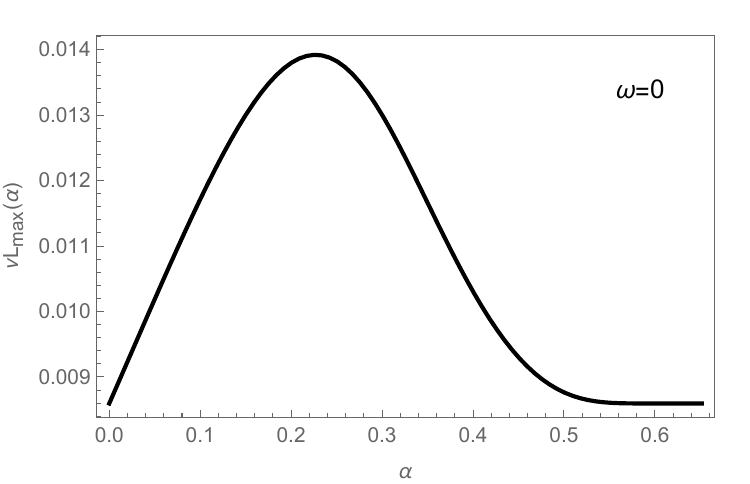} 
        \caption{ $\omega=0$}
        \label{8b}
    \end{subfigure}

    \vspace{0.2cm} 

    \begin{subfigure}[t]{0.4\textwidth}
        \centering
        \includegraphics[width=\textwidth]{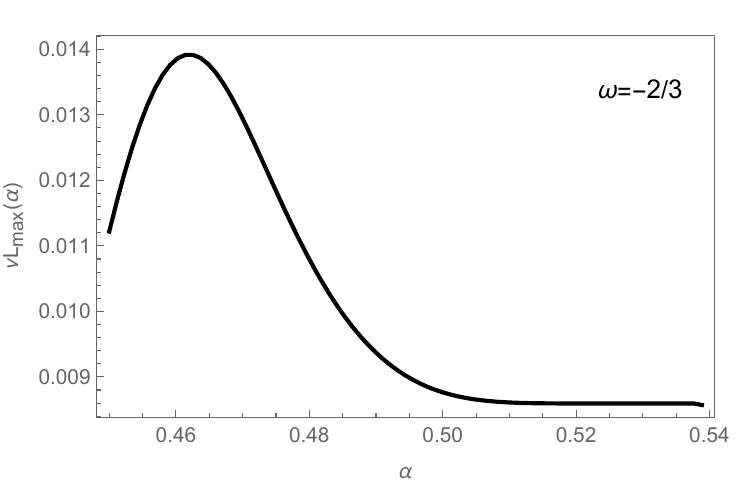} 
        \caption{ $\omega=-2/3$}
        \label{8c}
    \end{subfigure}
    \hspace{0.75cm} 
    \begin{subfigure}[t]{0.4\textwidth}
        \centering
        \includegraphics[width=\textwidth]{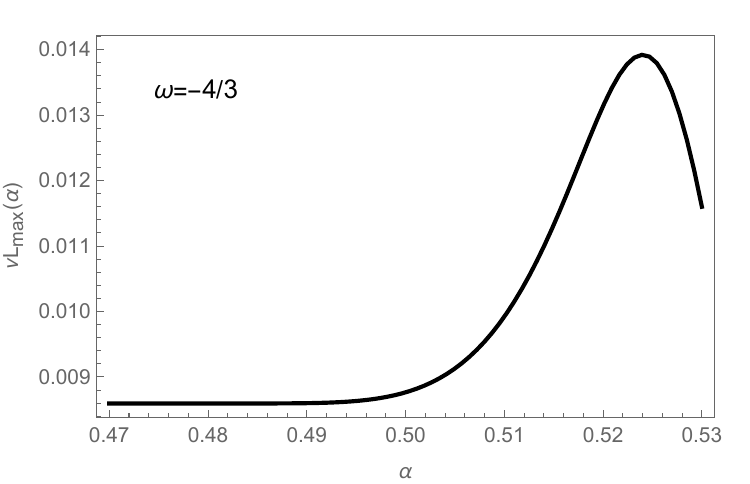} 
        \caption{ $\omega=-4/3$}
        \label{8d}
    \end{subfigure}

    \caption{The variation curves of the maximum of the emission spectrum ${\nu L}_\text{\rm{max}}(\alpha)$ for different $\omega$.}
    \label{8}
\end{figure}

Figures \ref{7} and \ref{8} depict the spectral energy distributions of accretion disks around a QFMGBH and the variation of the maximum values of the emission spectrum as \( \alpha \) changes. From figures \ref{7}(a) and \ref{8}(a), we observe that as \( \alpha \) increases, the emission spectrum for \( \omega = 1/3 \) initially shows a period of insignificant increase before decreasing. From figures \ref{7} and \ref{8}, the variation in the emission spectrum for \( \omega = 0 \), \( \omega = -2/3 \), and \( \omega = -4/3 \) follows a trend similar to that of the energy flux and the radiation temperature. Additionally, we find that the spectral energy distribution of the accretion disk around a QFMGBH is higher than that around a Schwarzschild black hole for different values of \( \omega \). Moreover, the peak values of the emission spectrum for different \( \omega \) all converge to a common limit of \( 0.0086 \).

From the above discussion, it can be inferred that the energy flux, the temperature, and the emission spectrum are higher from the QFMGBH as compared to the Schwarzschild BH. This is because that the $r_{\rm{ISCO}}$ of QFMGBH is smaller than or equal to that of Schwarzschild BH, so the disk around a QFMGBH is closer to the BH.

\subsection{The radiative efficiency} 

Now, we investigate the radiative efficiency \( \epsilon \). During mass accretion around a QFMGBH, the radiative efficiency is a key parameter that characterizes the central object's ability to convert its rest mass into outward radiating energy. We can define the radiative efficiency as the ratio of two rates: the rate at which radiation energy escapes from the disk surface to infinity, and the rate of mass energy transfer of the central dense object during mass accretion \cite{Novikov:1973kta, Page:1974he}. We assume that all photons can escape to infinity, so the radiative efficiency depends solely on the specific energy of the particle at the innermost stable circular orbit, \( r_{\rm{ISCO}} \) \cite{Novikov:1973kta, Page:1974he}.

\begin{equation}
    \epsilon=1-E_{\rm{ISCO}}
\end{equation}

\begin{figure}[htbp]
    \centering
    \begin{subfigure}[t]{0.4\textwidth}
        \centering
        \includegraphics[width=\textwidth]{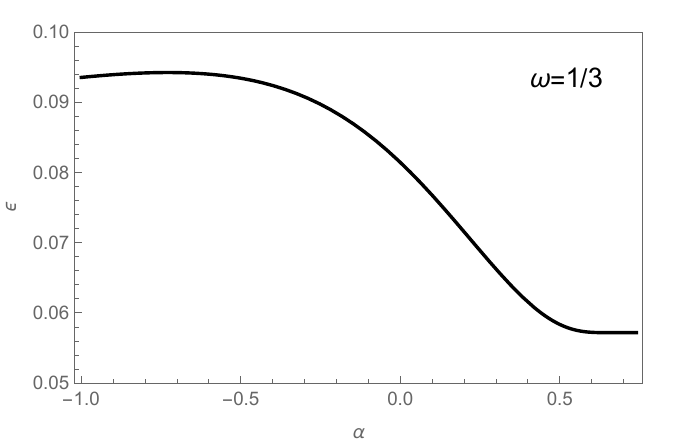} 
        \caption{ $\omega=1/3$}
        \label{9a}
    \end{subfigure}
    \hspace{0.75cm} 
    \begin{subfigure}[t]{0.4\textwidth}
        \centering
        \includegraphics[width=\textwidth]{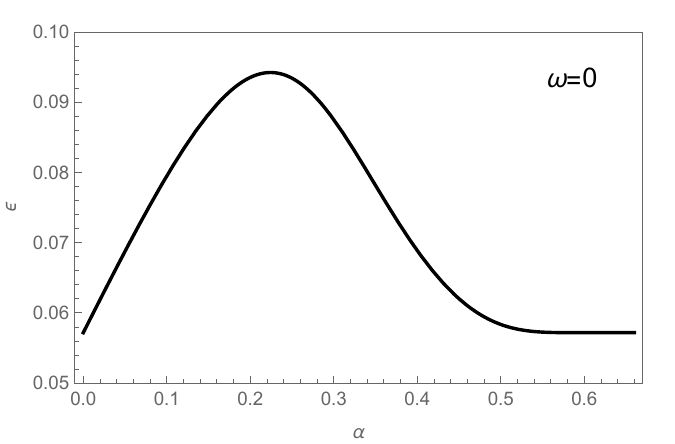} 
        \caption{ $\omega=0$}
        \label{9b}
    \end{subfigure}

    \vspace{0.2cm} 

    \begin{subfigure}[t]{0.4\textwidth}
        \centering
        \includegraphics[width=\textwidth]{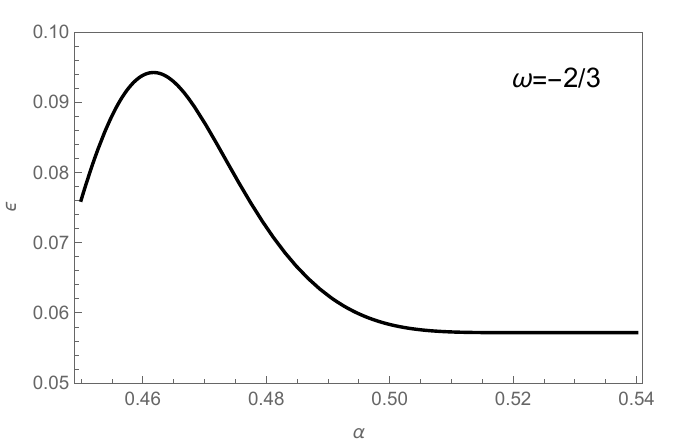} 
        \caption{ $\omega=-2/3$}
        \label{9c}
    \end{subfigure}
    \hspace{0.75cm} 
    \begin{subfigure}[t]{0.4\textwidth}
        \centering
        \includegraphics[width=\textwidth]{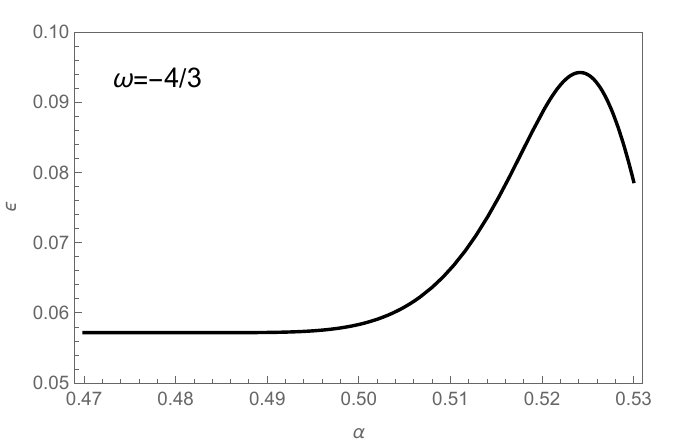} 
        \caption{ $\omega=-4/3$}
        \label{9d}
    \end{subfigure}

    \caption{The variation curves of the efficiency  $\epsilon$ for different $\omega$.} 
    \label{9}
\end{figure}

Figure \ref{9} shows the variation of the radiative efficiency for different values of \( \omega \). From figure \ref{9}, we can observe that the changes in radiative efficiency for different \( \omega \) are similar to those of the emission spectrum. Additionally, the radiative efficiency of the disk around a QFMGBH for different \( \omega \) is higher than that around a Schwarzschild BH. Moreover, the radiative efficiency for different \( \omega \) approaches a limit value of \( 0.057 \).

\subsection{Some important data} 

Table \ref{tab1} displays the maximum values of the energy flux \( F_\text{max}(r) \), the radiation temperature \( T_\text{max}(r) \), and the emission spectrum \( \nu L(\nu)_\text{max} \) for an accretion disk around a QFMGBH. It also provides the ISCO radius \( r_{\rm{ISCO}} \), the critical cutoff frequency \( \nu_\text{crit} \), and the radiative efficiency \( \epsilon \).

\begin{table}[htbp]
   \centering
    \begin{tabular}{c c c c c c c c}
        \hline
        \hline
         $\omega$&  $\alpha$&  $r_{\rm{ISCO}}$&  $F_{\rm{max}}$&  $T_{\rm{max}}$&  $\nu_{\rm{crit}}$&  $\nu L(\nu)_{\rm{max}}$& $\epsilon$\\\hline
         \hline
          GR &  $-$&  $6.00$&  $1.37\times10^{-5}$&  $0.095$&  $0.033$&  $0.0086$& $0.057$\\\hline
          \hline
         $1/3$&  $-0.8$&  $3.09$&  $8.01\times10^{-5}$&  $0.148$&  $0.048$&  $0.0139$& $0.091$\\\hline
         &  $-0.4$&  $3.33$&  $6.83\times10^{-5}$&  $0.143$&  $0.046$&  $0.0137$& $0.089$\\\hline
         & $ 0.0$&  $4.00$&  $4.25\times10^{-5}$&  $0.127$&  $0.042$&  $0.0121$& $0.078$\\\hline
         & $ 0.4$&  $5.49$&  $1.74\times10^{-5}$&  $0.101$&  $0.035$&  $0.0092$& $0.058$\\\hline\hline
         $0$&  $0.1$&  $2.93$&  $7.48\times10^{-5}$&  $0.146$&  $0.048$&  $0.0117$& $0.076$\\\hline
         &  $0.2$&  $3.04$&  $8.22\times10^{-5}$&  $0.149$&  $0.048$&  $0.0138$& $0.090$\\\hline
         &  $0.3$&  $3.64$&  $5.48\times10^{-5}$&  $0.135$&  $0.044$&  $0.0130$& $0.084$\\\hline
         &  $0.4$&  $4.85$&  $2.48\times10^{-5}$&  $0.111$&  $0.037$&  $0.0103$& $0.066$\\\hline\hline
         $-2/3$&  $0.45$&  $2.94$&  $7.17\times10^{-5}$&  $0.144$&  $0.048$&  $0.0112$& $0.073$\\\hline
 & $0.46$& $3.05$& $8.16\times10^{-5}$& $0.149$& $0.048$& $0.0138$&$0.090$\\\hline
 & $0.47$& $3.66$& $5.38\times10^{-5}$& $0.134$& $0.044$& $0.0129$&$0.084$\\\hline
 & $0.48$& $4.60$& $2.88\times10^{-5}$& $0.115$& $0.039$& $0.0108$&$0.069$\\\hline\hline
$ -4/3$& $0.515$& $4.35$& $3.37\times10^{-5}$& $0.120$& $0.040$& $0.0113$&$0.073$\\\hline
 & $0.520$& $3.58$& $5.70\times10^{-5}$& $0.136$& $0.044$& $0.0131$&$0.085$\\\hline
 & $0.525$& $3.06$& $8.14\times10^{-5}$& $0.149$& $0.048$& $0.0139$&$0.090$\\\hline
 & $0.530$& $2.93$& $7.41\times10^{-5}$& $0.146$& $0.048$& $0.0116$&$0.075$\\\hline\hline
    \end{tabular}
    \caption{The maximum values of the flux \( F_\text{max}(r) \), the radiation temperature \( T_\text{max}(r) \), the emission spectrum \( \nu L(\nu)_\text{max} \) of an accretion disk around a QFMGBH, the ISCO radius \( r_{\rm{ISCO}} \), the critical cutoff frequency \( \nu_\text{crit} \), and the radiative efficiency \( \epsilon \) are presented.}
    \label{tab1}
\end{table}

\subsection{The shadow}

In this section, we focus on the shadow images of QFMGBHs. The BH is surrounded by various objects (e.g., the accretion disk) that emit photons, which can be received by an observer at infinity, providing information about the BH and the surrounding matter. To obtain shadow images of QFMGBHs, we use an open-source general relativistic ray-tracing code called GYOTO \cite{Vincent_2011, aimar2023gyoto20polarizedrelativistic}. The original GYOTO code for the thin accretion disk, which is geometrically thin but optically thick, describes the Kerr spacetime. We modify the original code to simulate images of the geometrically thin and optically thick accretion disk around QFMGBHs.
\begin{figure}[htbp]
    \centering
    \begin{subfigure}[t]{0.18\textwidth}
        \centering
        \includegraphics[width=\textwidth]{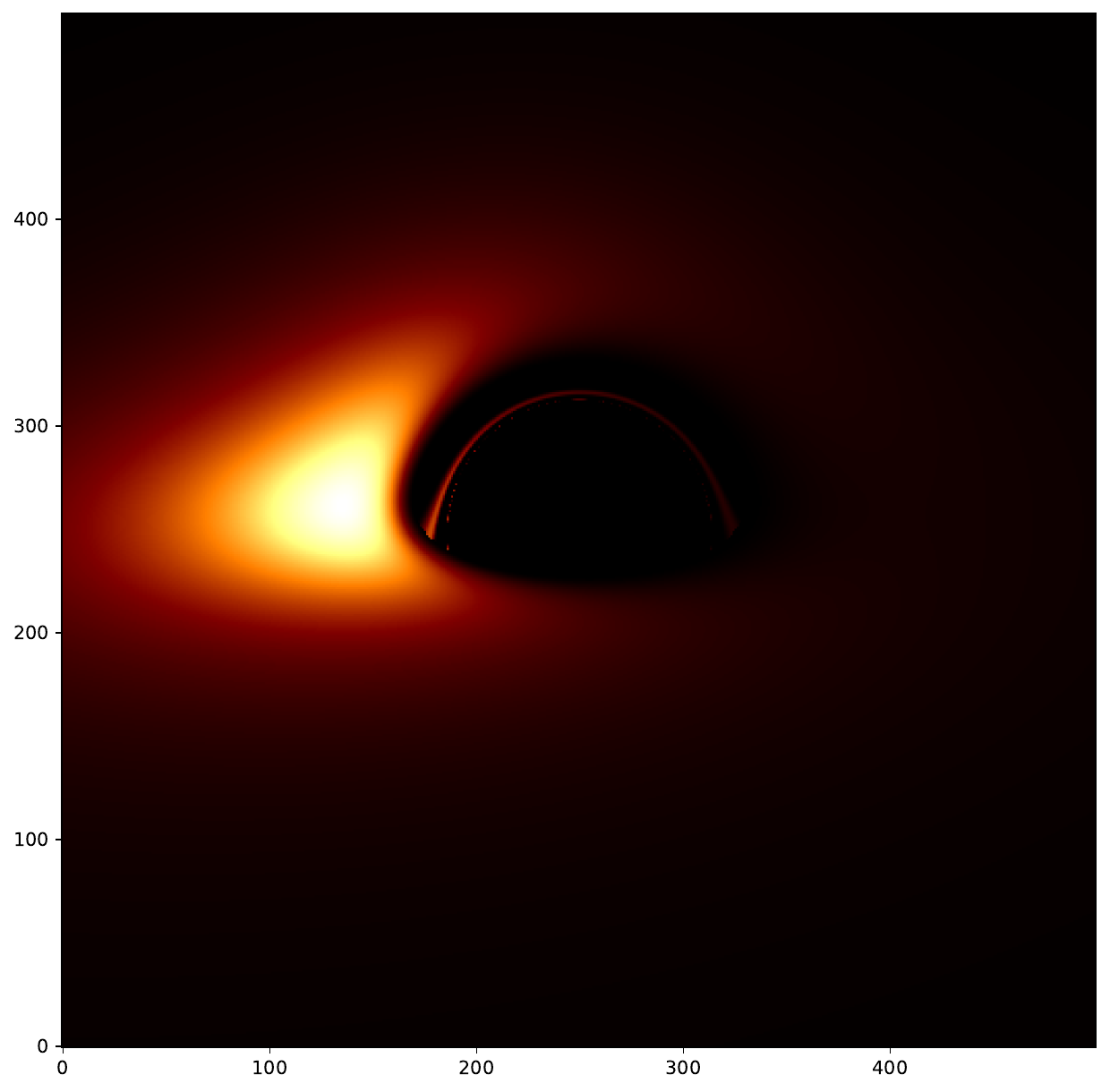} 
        \caption{$\alpha=-0.8$}
        \label{10a}
    \end{subfigure}
    \hspace{0.001cm} 
    \begin{subfigure}[t]{0.18\textwidth}
        \centering
        \includegraphics[width=\textwidth]{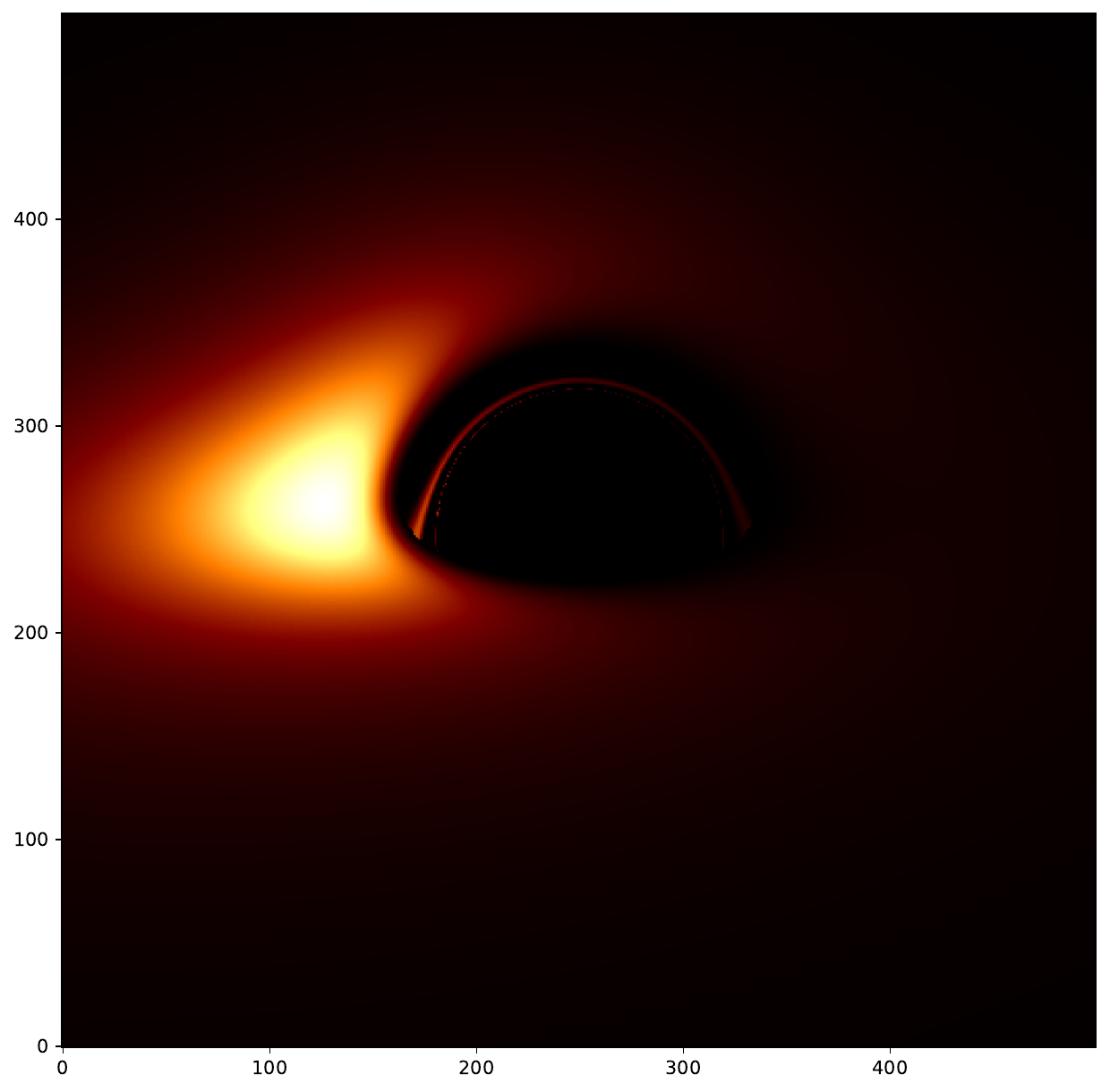} 
        \caption{$\alpha=-0.4$}
        \label{10b}
    \end{subfigure}
     \hspace{0.001cm} 
    \begin{subfigure}[t]{0.18\textwidth}
        \centering
        \includegraphics[width=\textwidth]{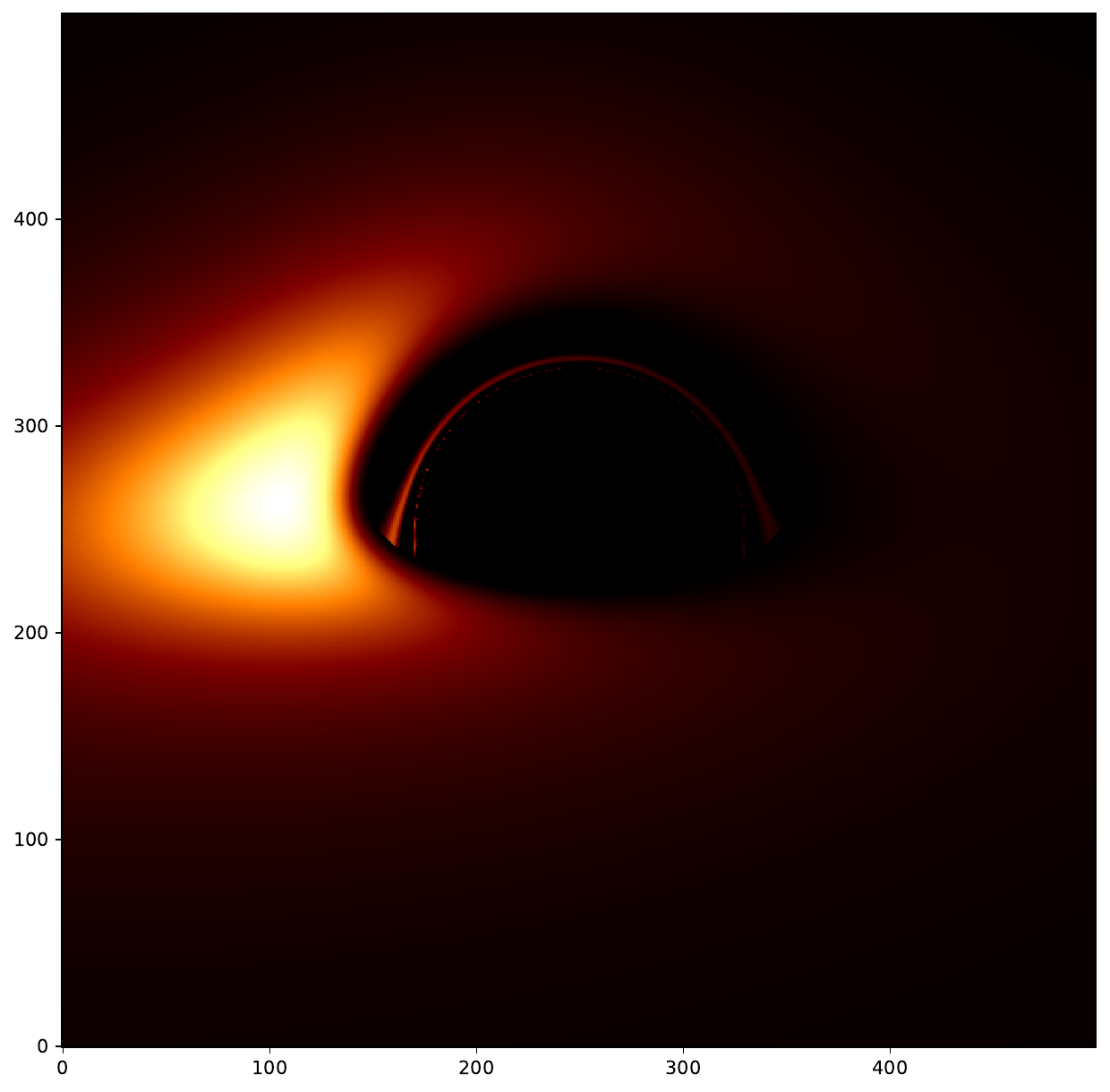} 
        \caption{$\alpha=0.0$}
        \label{10c}
    \end{subfigure}
    \hspace{0.001cm} 
    \begin{subfigure}[t]{0.18\textwidth}
        \centering
        \includegraphics[width=\textwidth]{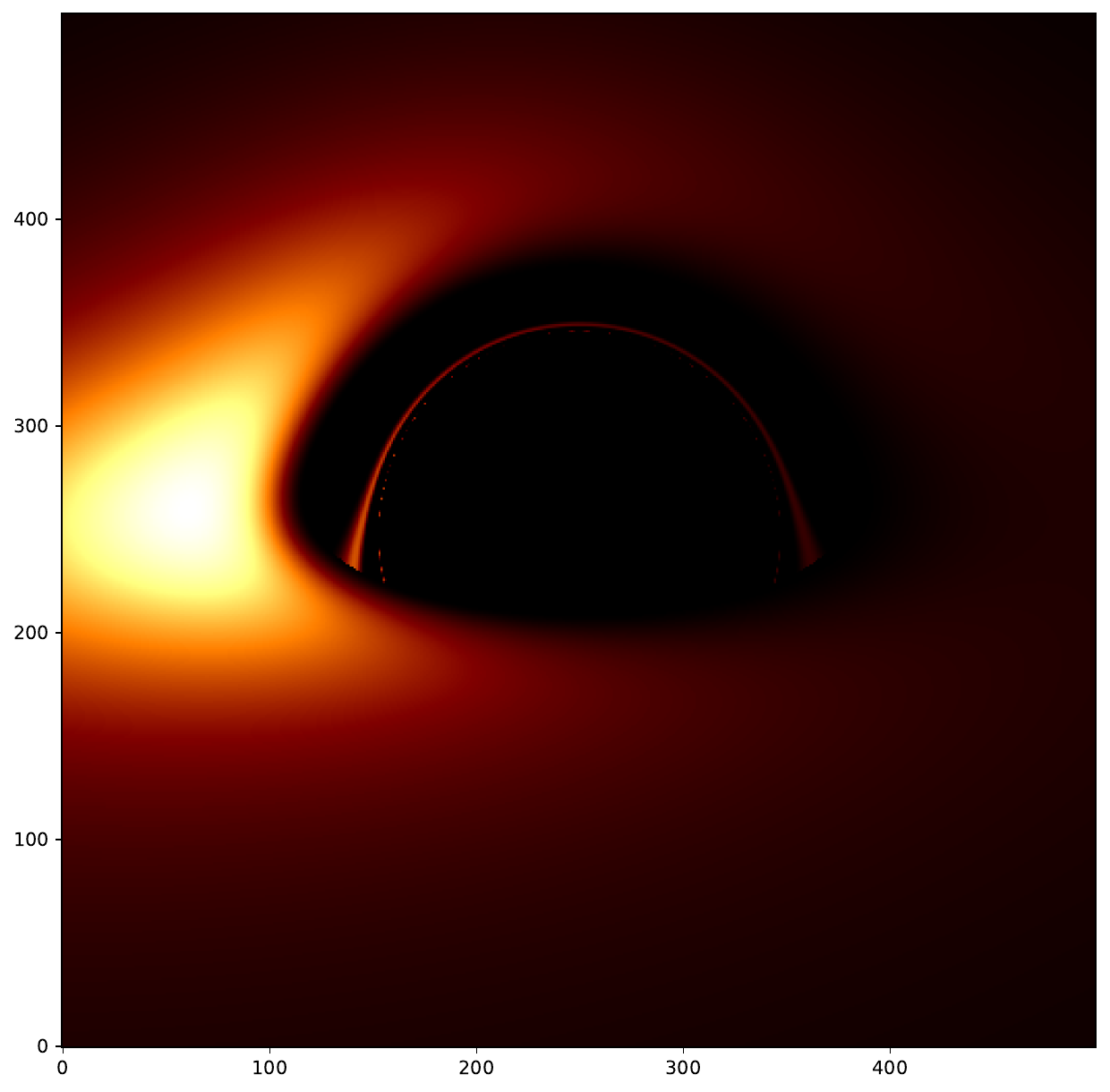} 
        \caption{$\alpha=0.4$}
        \label{10d}
  
    \end{subfigure}
 \hspace{0.001cm} 
    \begin{subfigure}[t]{0.18\textwidth}
    
        \centering
        \includegraphics[width=\textwidth]{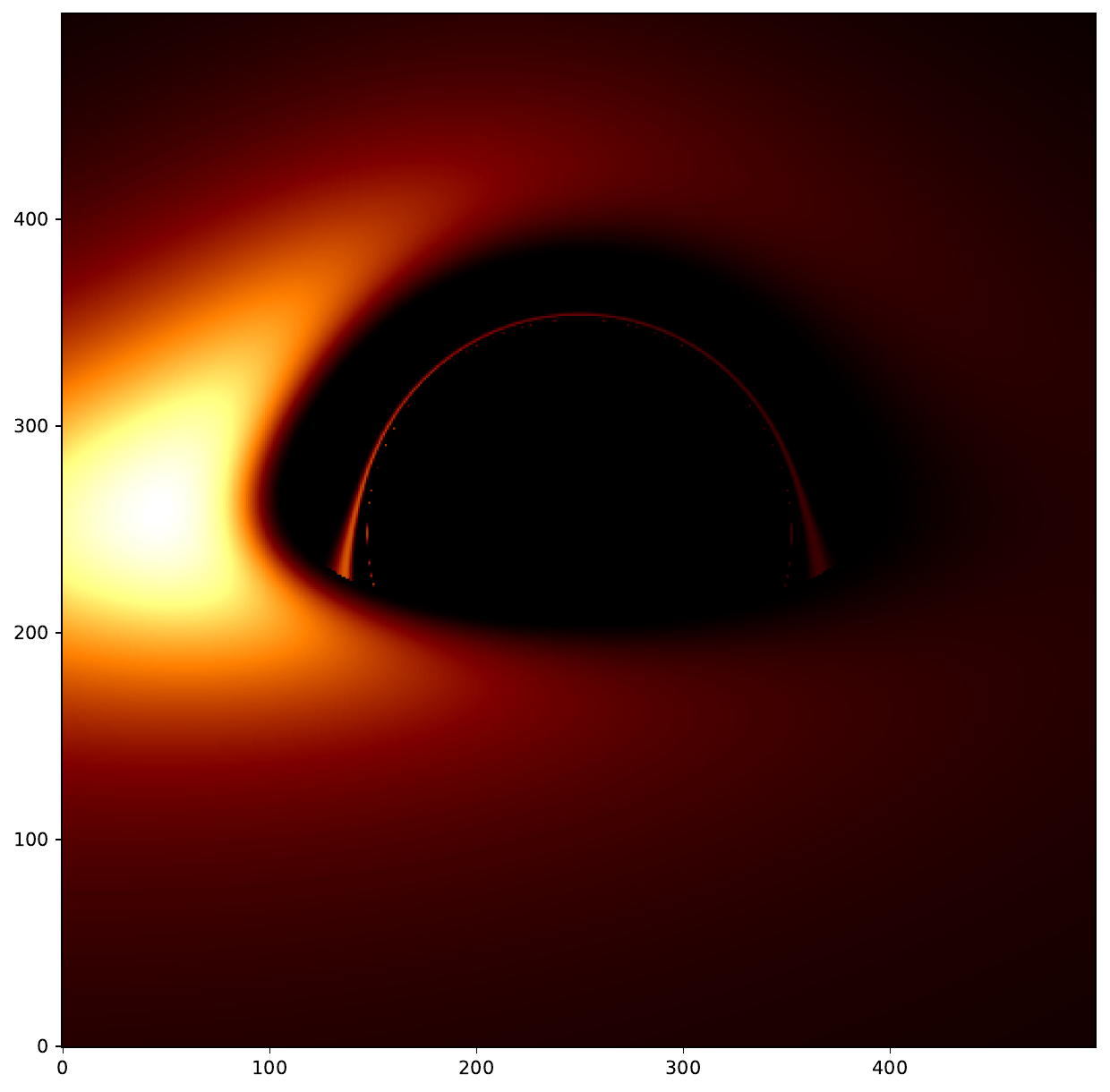} 
        \caption{Schwarzschild}
        \label{10e}
    \end{subfigure}
  
    \caption{The accretion disks generated using GYOTO code for $\omega=1/3$.}
    \label{10}
\end{figure}

\begin{figure}[htbp]
    \centering
    \begin{subfigure}[t]{0.18\textwidth}
        \centering
        \includegraphics[width=\textwidth]{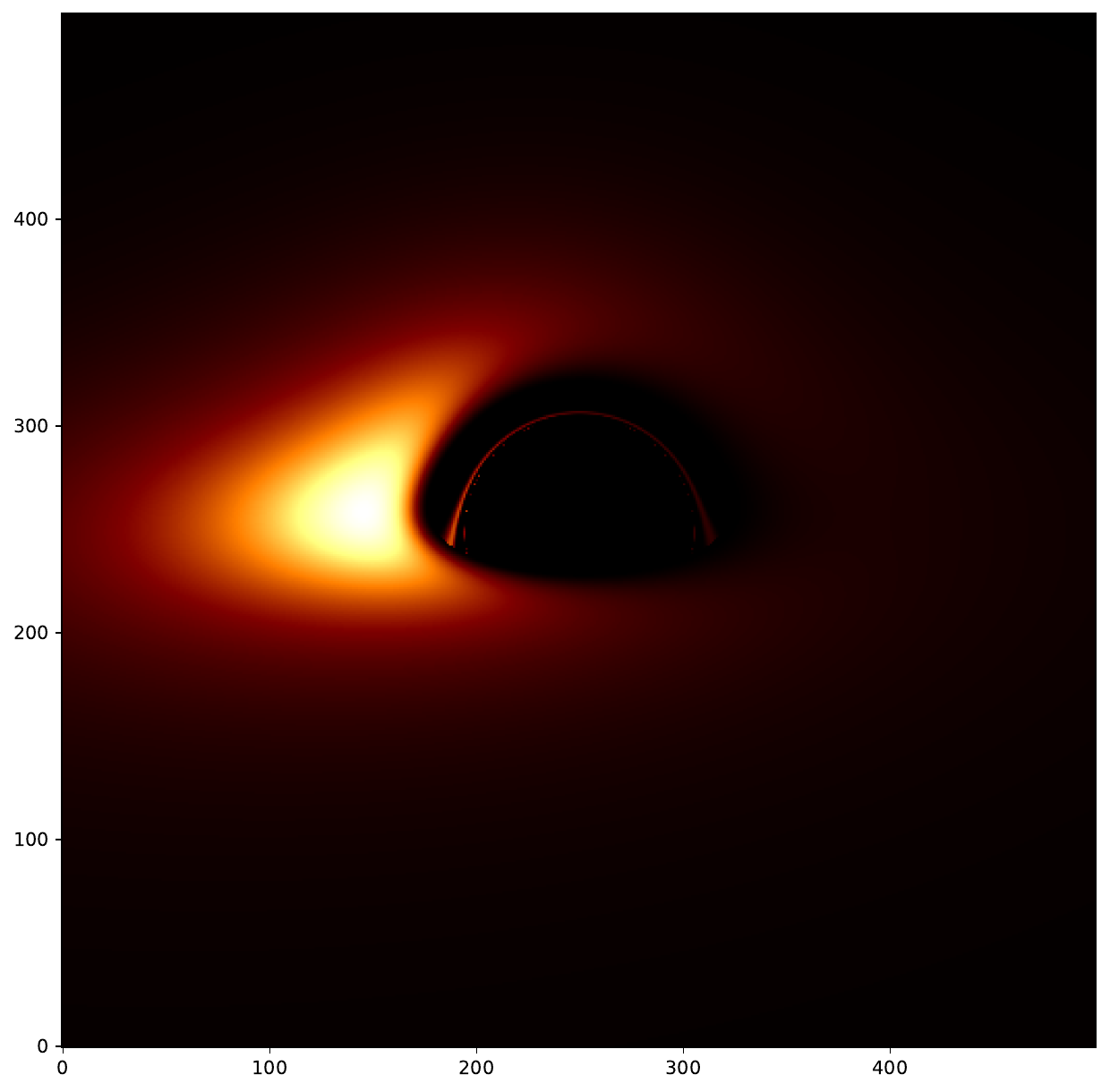} 
        \caption{$\alpha=0.1$}
        \label{11a}
    \end{subfigure}
    \hspace{0.001cm} 
    \begin{subfigure}[t]{0.18\textwidth}
        \centering
        \includegraphics[width=\textwidth]{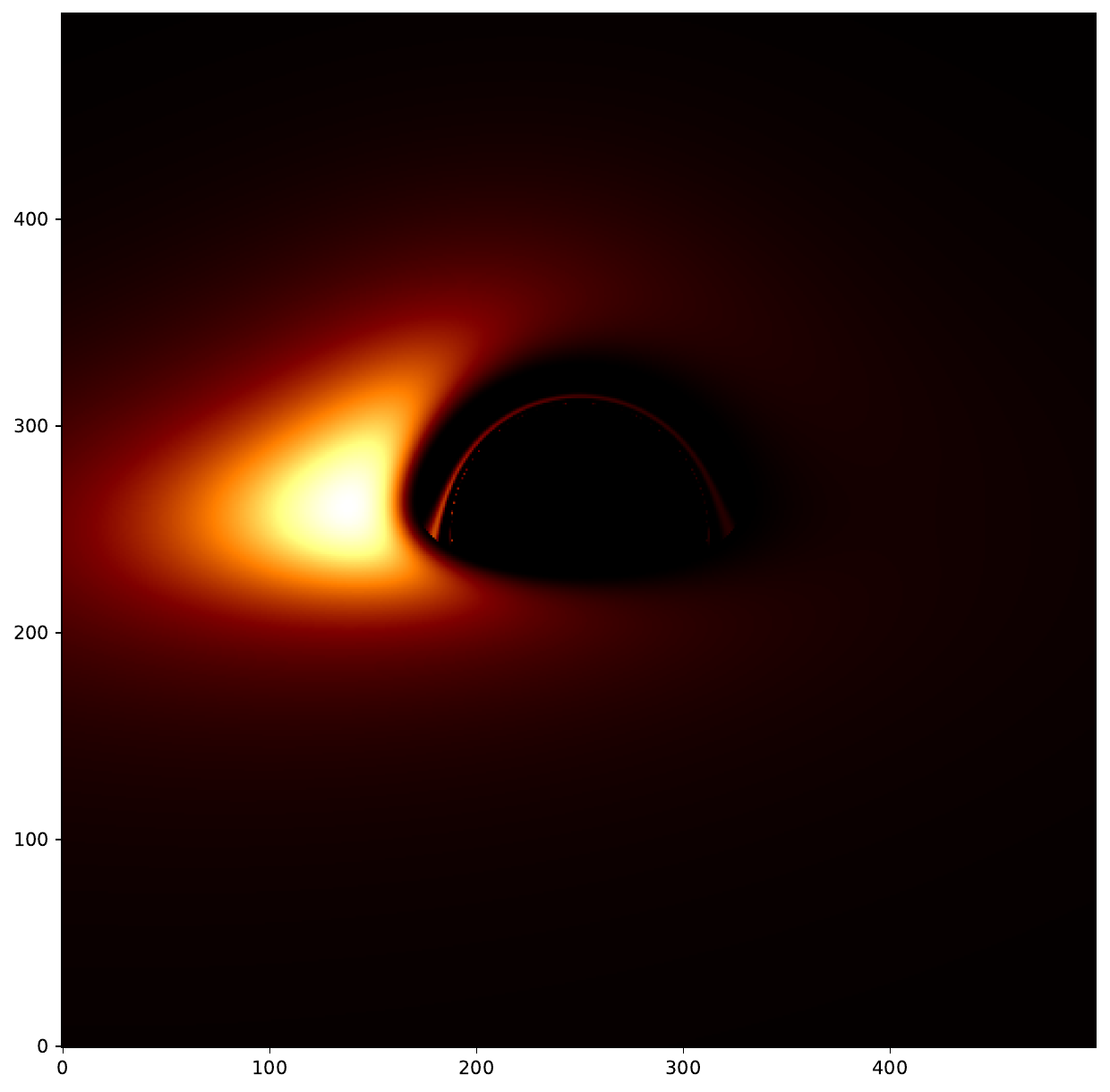} 
        \caption{$\alpha=0.2$}
        \label{11b}
    \end{subfigure}
     \hspace{0.001cm} 
    \begin{subfigure}[t]{0.18\textwidth}
        \centering
        \includegraphics[width=\textwidth]{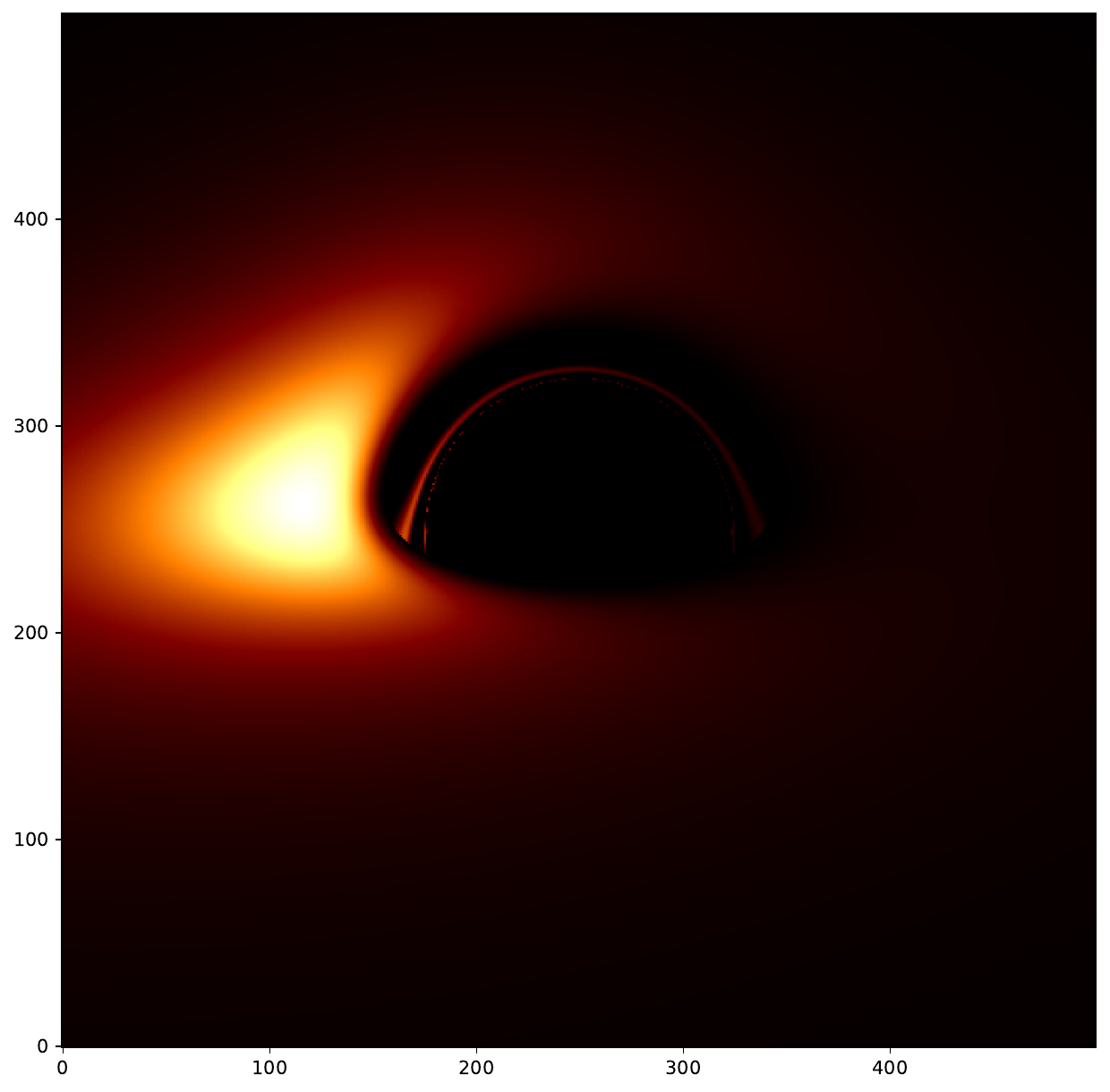} 
        \caption{$\alpha=0.3$}
        \label{11c}
    \end{subfigure}
    \hspace{0.001cm} 
    \begin{subfigure}[t]{0.18\textwidth}
        \centering
        \includegraphics[width=\textwidth]{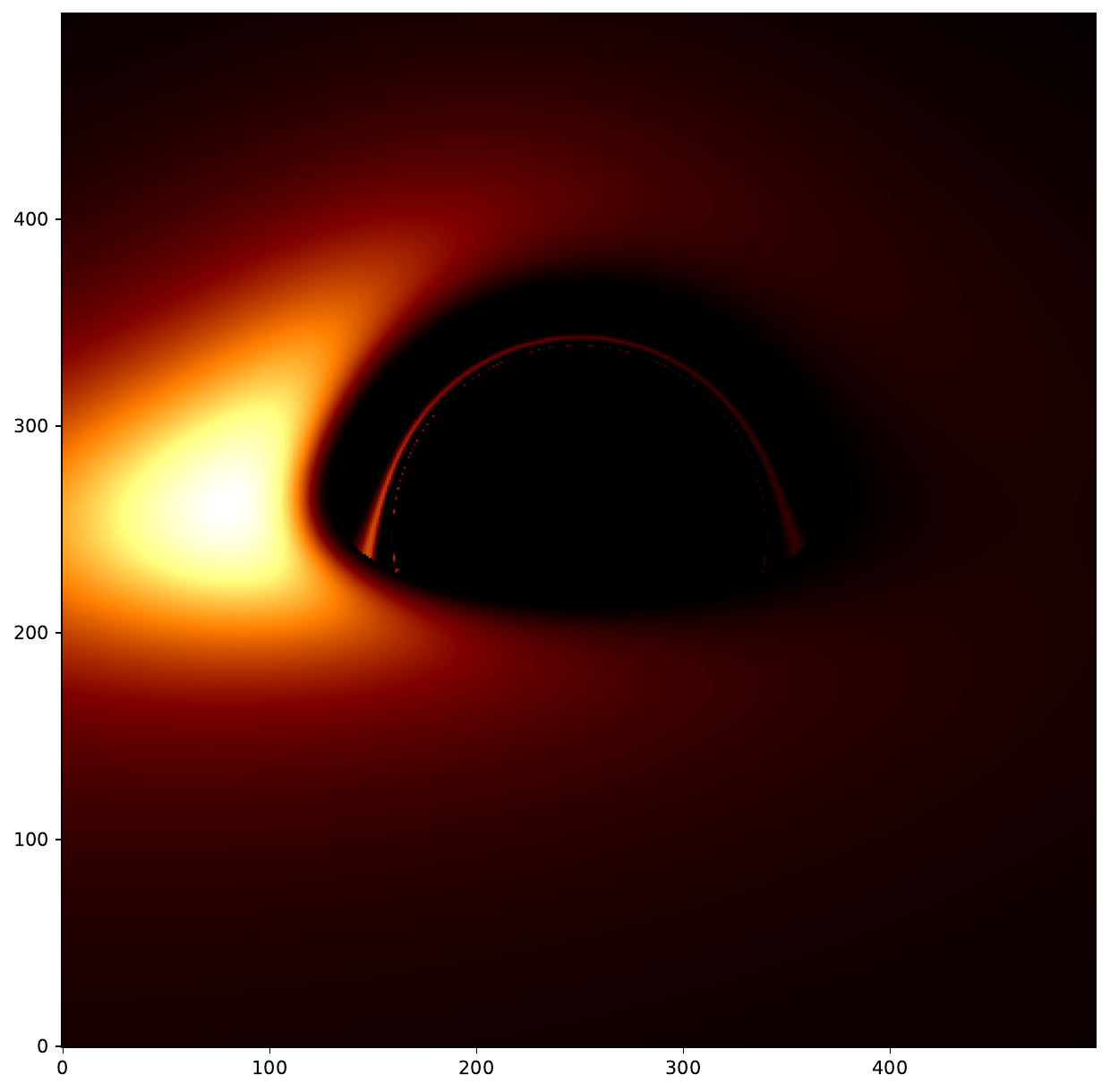} 
        \caption{$\alpha=0.4$}
        \label{11d}
    \end{subfigure}
\hspace{0.001cm} 
    \begin{subfigure}[t]{0.18\textwidth}
        \centering
        \includegraphics[width=\textwidth]{PageThorneGR.pdf} 
        \caption{Schwarzschild}
        \label{11e}
    \end{subfigure}
    \caption{The accretion disks generated using GYOTO code for $\omega=0$.}
    \label{11}
\end{figure}

\begin{figure}[htbp]
    \centering
    \begin{subfigure}[t]{0.18\textwidth}
        \centering
        \includegraphics[width=\textwidth]{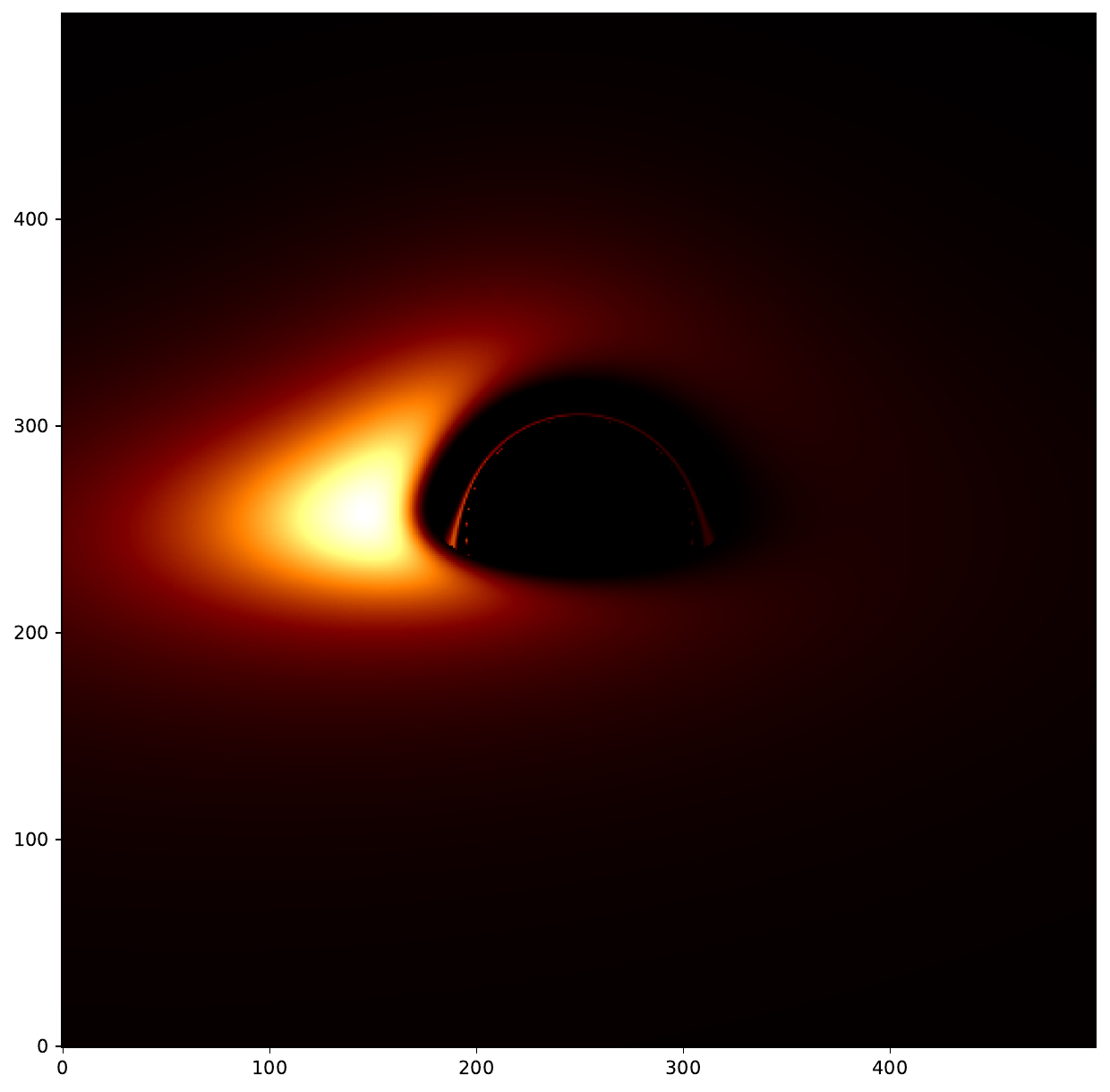} 
        \caption{$\alpha=0.45$}
        \label{12a}
    \end{subfigure}
    \hspace{0.001cm} 
    \begin{subfigure}[t]{0.18\textwidth}
        \centering
        \includegraphics[width=\textwidth]{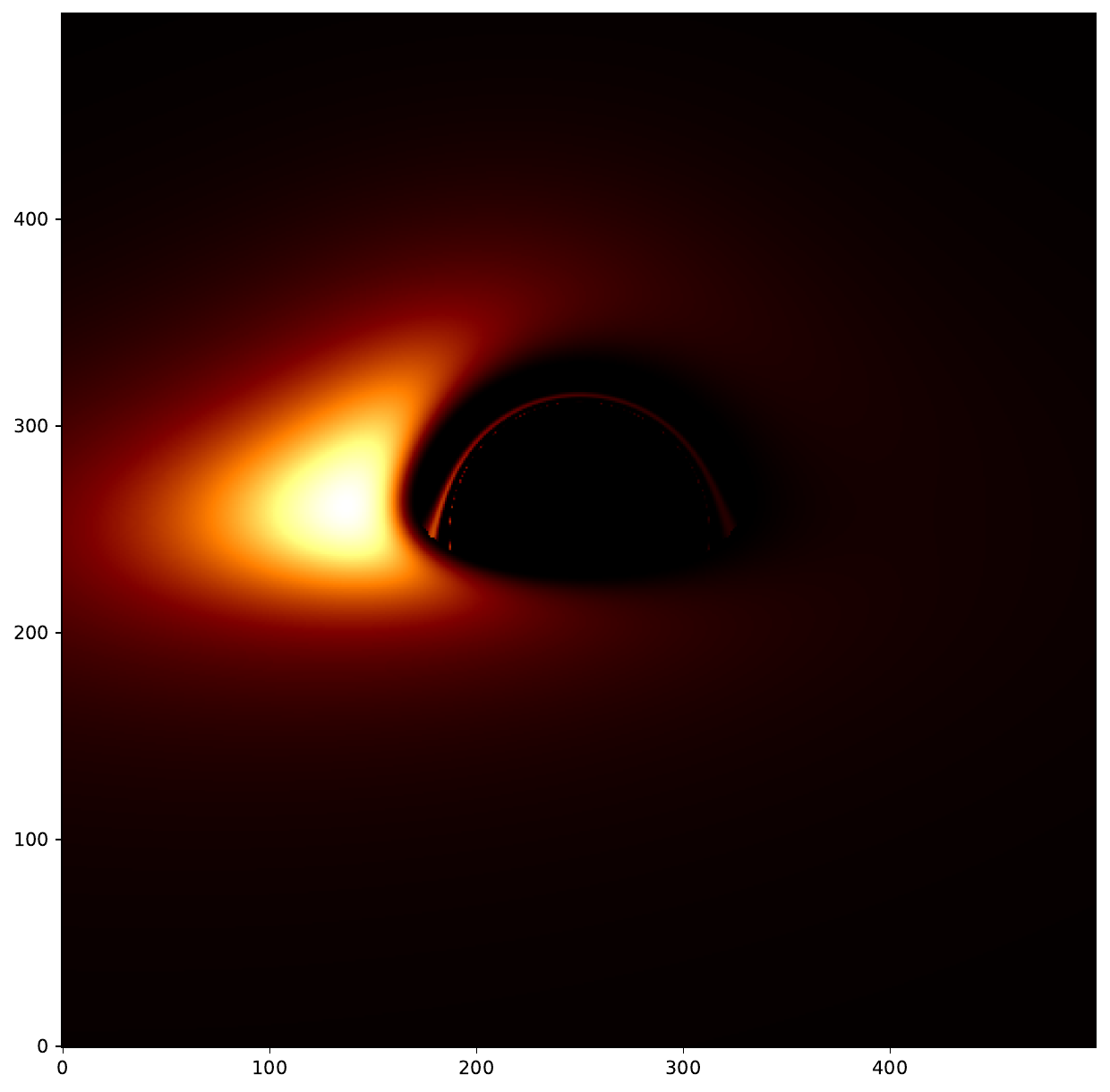} 
        \caption{$\alpha=0.46$}
        \label{12b}
    \end{subfigure}
     \hspace{0.001cm} 
    \begin{subfigure}[t]{0.18\textwidth}
        \centering
        \includegraphics[width=\textwidth]{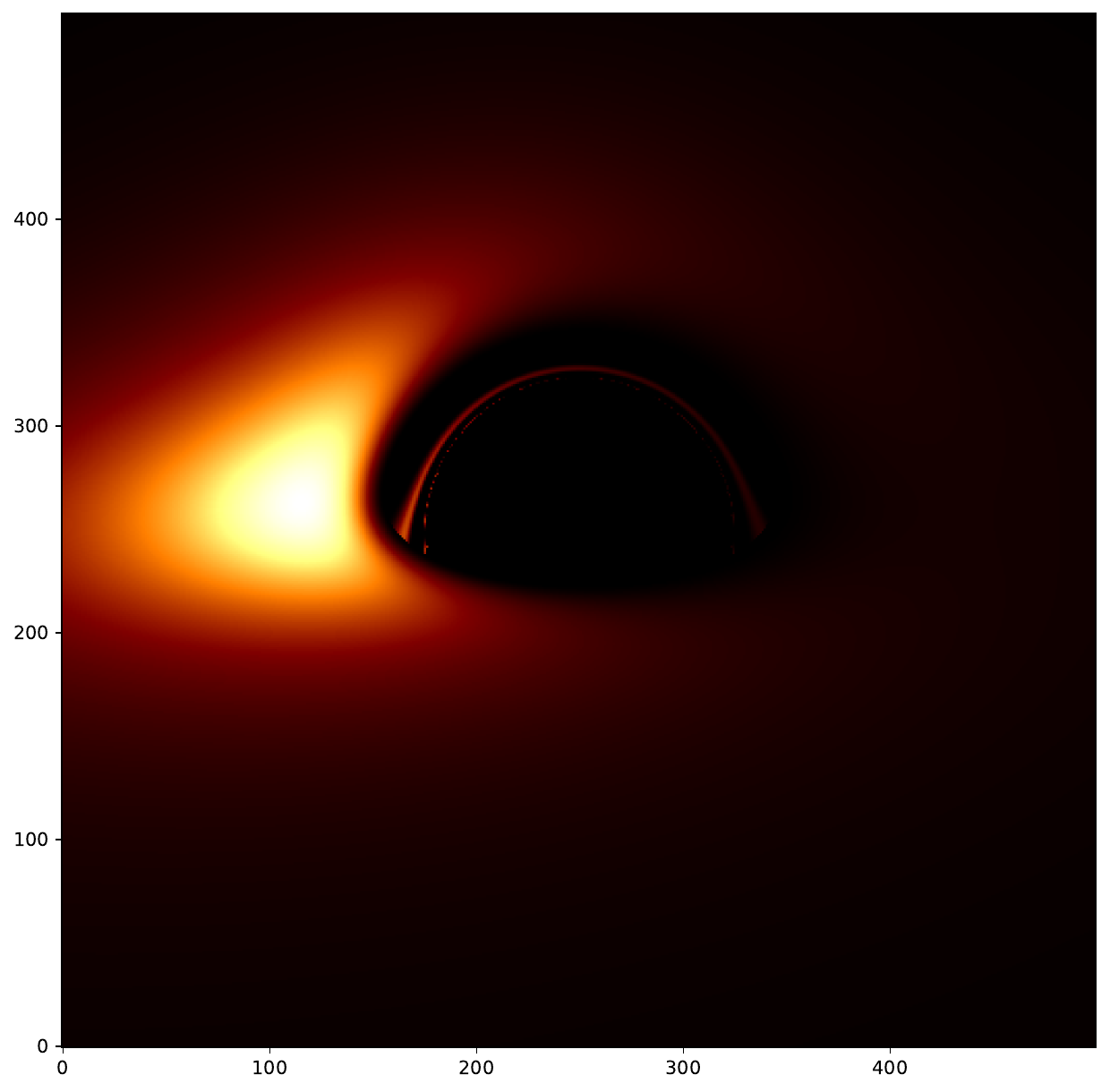} 
        \caption{$\alpha=0.47$}
        \label{12c}
    \end{subfigure}
    \hspace{0.001cm} 
    \begin{subfigure}[t]{0.18\textwidth}
        \centering
        \includegraphics[width=\textwidth]{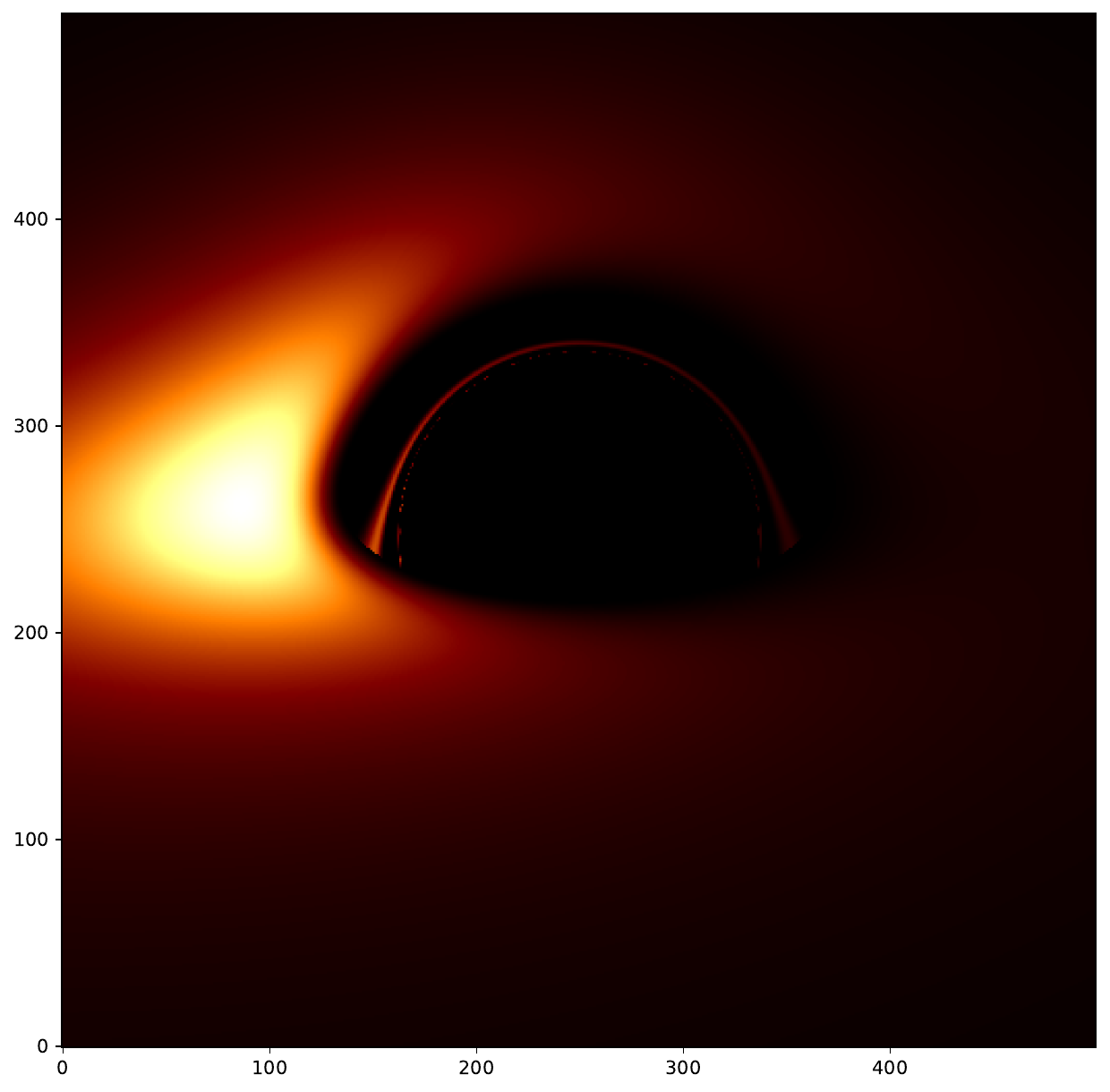} 
        \caption{$\alpha=0.48$}
        \label{12d}
    \end{subfigure}
\hspace{0.001cm} 
    \begin{subfigure}[t]{0.18\textwidth}
        \centering
        \includegraphics[width=\textwidth]{PageThorneGR.pdf} 
        \caption{Schwarzschild}
        \label{12e}
    \end{subfigure}

    \caption{The accretion disks generated using GYOTO code for $\omega=-2/3$.}
    \label{12}
\end{figure}
\begin{figure}[htbp]
    \centering
    \begin{subfigure}[t]{0.18\textwidth}
        \centering
        \includegraphics[width=\textwidth]{PageThorneGR.pdf} 
        \caption{Schwarzschild}
        \label{13a}
    \end{subfigure}
    \hspace{0.001cm} 
    \begin{subfigure}[t]{0.18\textwidth}
        \centering
        \includegraphics[width=\textwidth]{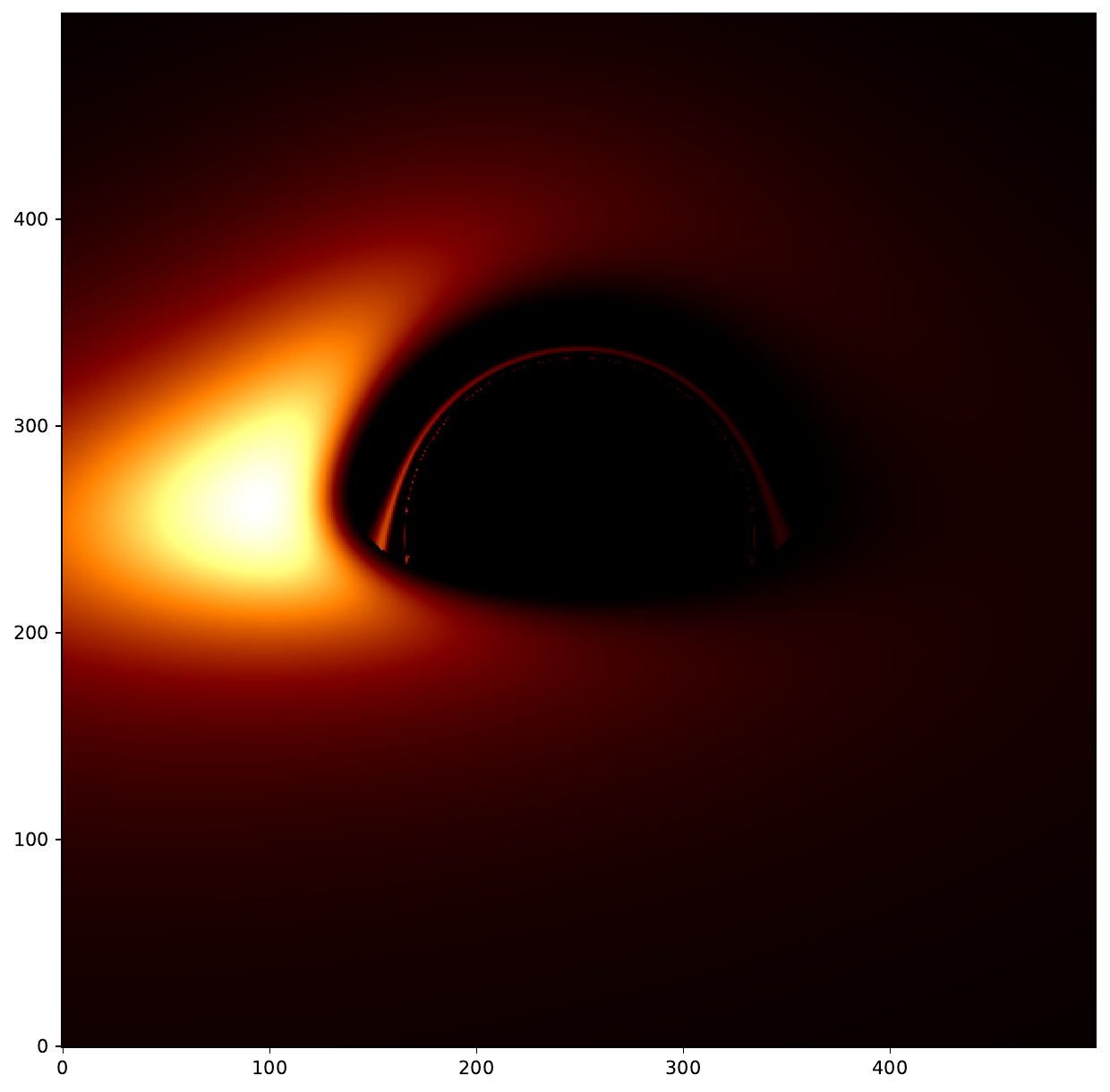} 
        \caption{$\alpha=0.515$}
        \label{13b}
    \end{subfigure}
     \hspace{0.001cm} 
    \begin{subfigure}[t]{0.18\textwidth}
        \centering
        \includegraphics[width=\textwidth]{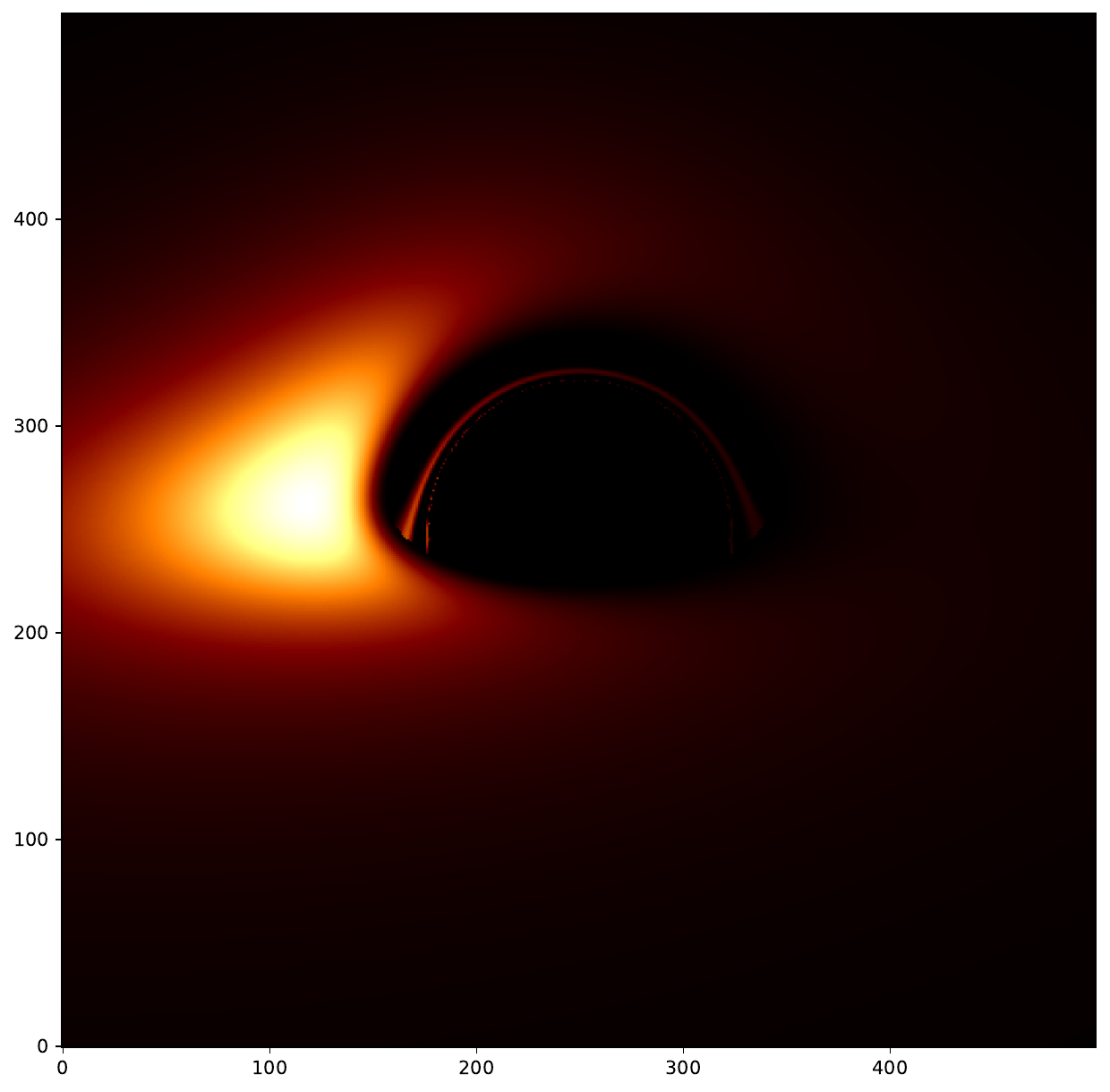} 
        \caption{$\alpha=0.520$}
        \label{13c}
    \end{subfigure}
    \hspace{0.001cm} 
    \begin{subfigure}[t]{0.18\textwidth}
        \centering
        \includegraphics[width=\textwidth]{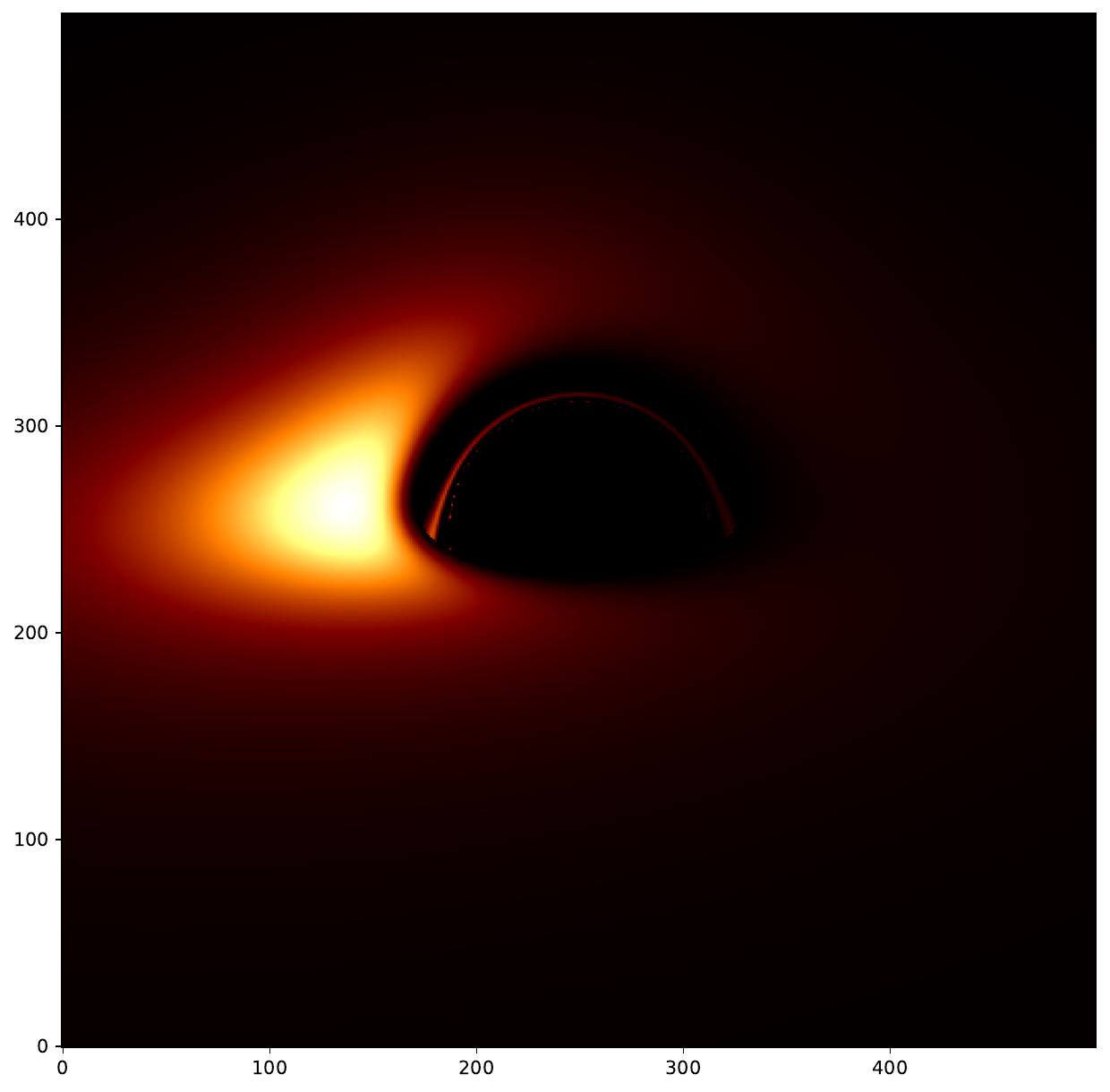} 
        \caption{$\alpha=0.525$}
        \label{13d}
    \end{subfigure}
 \hspace{0.001cm} 
    \begin{subfigure}[t]{0.18\textwidth}
        \centering
        \includegraphics[width=\textwidth]{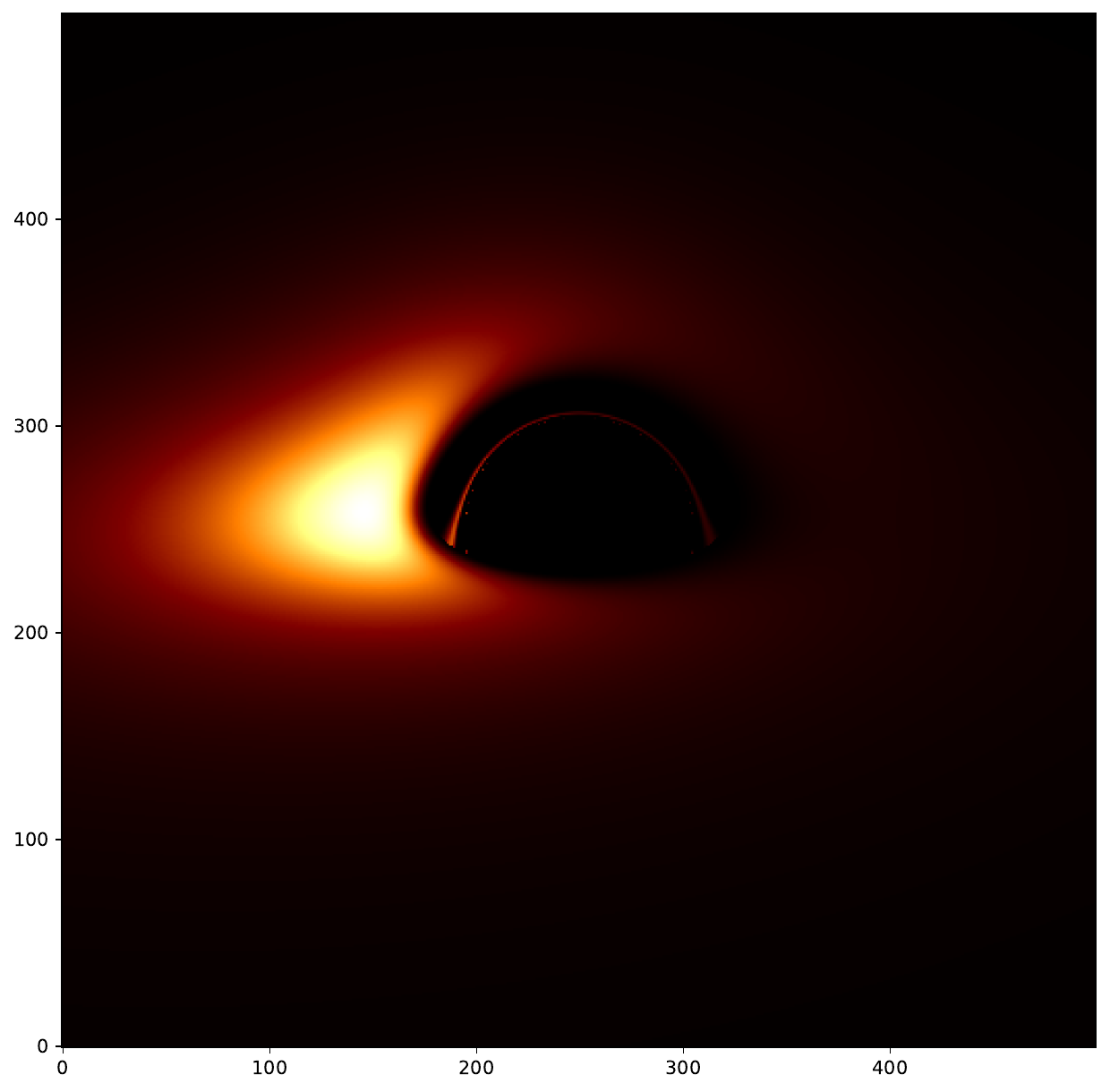} 
        \caption{$\alpha=0.530$}
        \label{13e}
    \end{subfigure}
    
    \caption{The accretion disks generated using GYOTO code for $\omega=-4/3$.}
    \label{13}
\end{figure}

In figures \ref{10}, \ref{11}, \ref{12}, and \ref{13}, we present the accretion disks obtained using GYOTO, with the observation radius set to \( r_{\rm o} = 80M \), the observation angle \( \theta_{\rm o} = 70^\circ \). The accretion disks in figures \ref{10}, \ref{11}, and \ref{12} correspond to fixed values of \( \omega = 1/3 \), \( \omega = 0 \), and \( \omega = -2/3 \), respectively. For comparison, we also present the shadow of the Schwarzschild BH (no fluctuations and no fluids around it). From figures \ref{10}, \ref{11}, and \ref{12}, we observe that as \( \alpha \) increases, the area of the shadow of the BH increases. Figure \ref{11} with \( \omega =0 \) show the shadow of Schwarzschild like BH with reduced mass $2M-D$. Figure \ref{13} corresponds to a fixed value of \( \omega = -4/3 \), and from which, we observe that as \( \alpha \) increases, the area of the shadow of the QFMGBH decreases. Furthermore, by comparing figures \ref{10}, \ref{11}, \ref{12}, and \ref{13}, we find that as \( \omega \) decreases, the absolute value of the rate of change of the area increases. Notably, for all four values of \( \omega \), the area of the shadow of the QFMGBH is smaller than that of the Schwarzschild BH.
 \begin{figure}[htbp]
    \centering
    \begin{subfigure}[t]{0.18\textwidth}
        \centering
        \includegraphics[width=\textwidth]{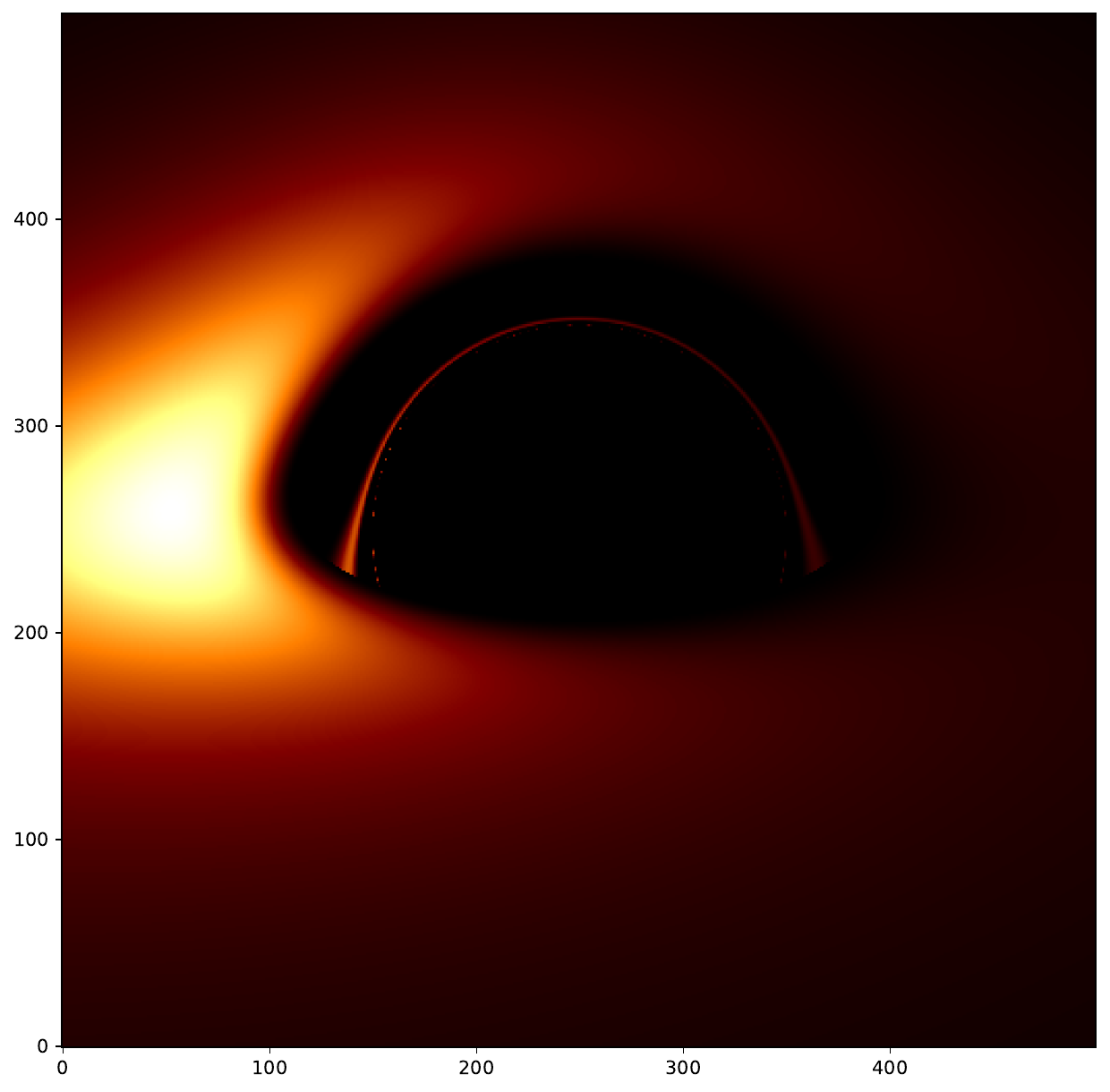} 
        \caption{$\omega=1/3$}
        \label{14a}
    \end{subfigure}
    \hspace{0.001cm} 
    \begin{subfigure}[t]{0.18\textwidth}
        \centering
        \includegraphics[width=\textwidth]{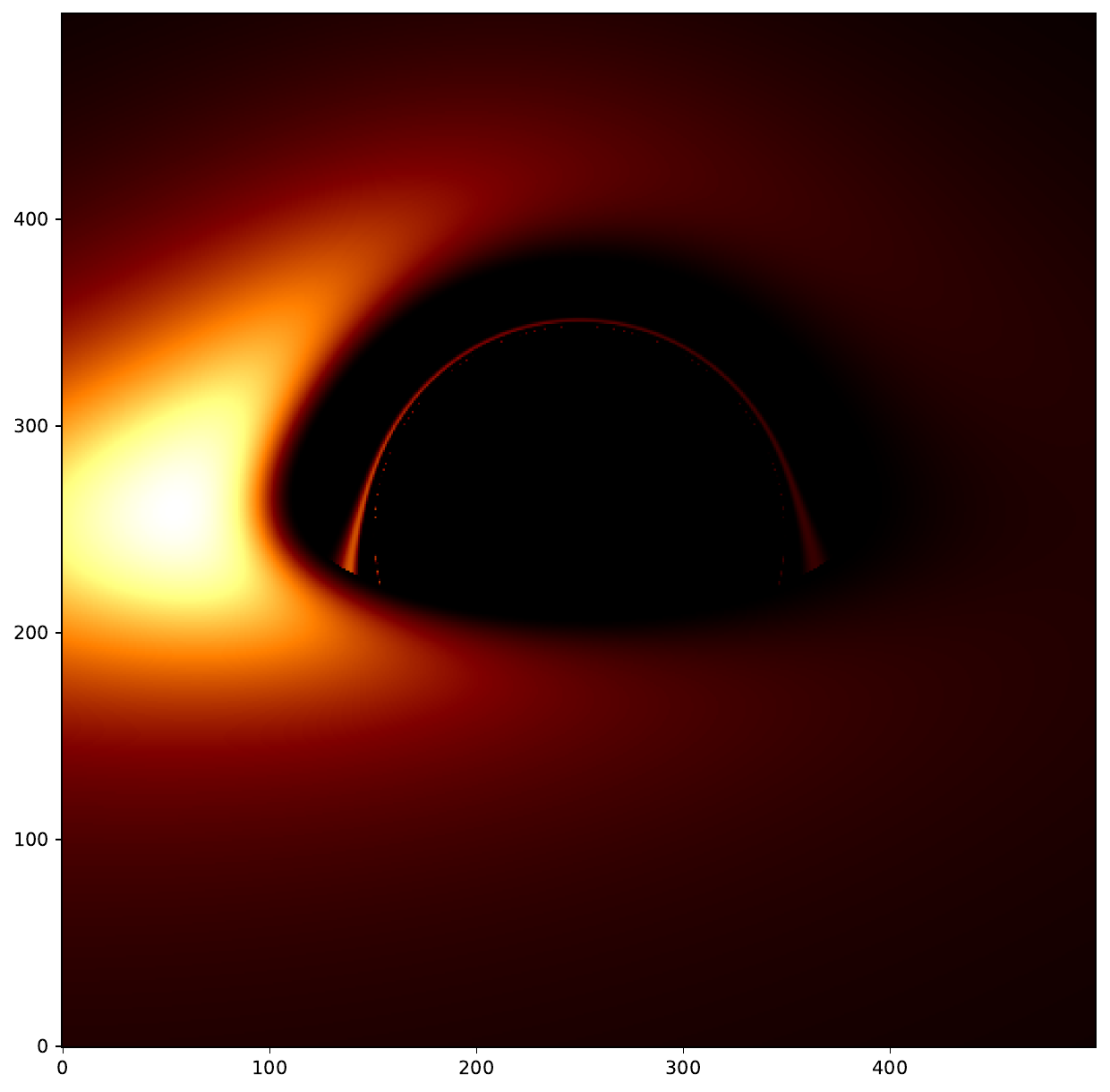} 
        \caption{$\omega=0$}
        \label{14b}
    \end{subfigure}
     \hspace{0.001cm} 
    \begin{subfigure}[t]{0.18\textwidth}
        \centering
        \includegraphics[width=\textwidth]{f0.66670.48-70c.pdf} 
        \caption{$\omega=-2/3$}
        \label{14c}
    \end{subfigure}
    \hspace{0.001cm} 
    \begin{subfigure}[t]{0.18\textwidth}
        \centering
        \includegraphics[width=\textwidth]{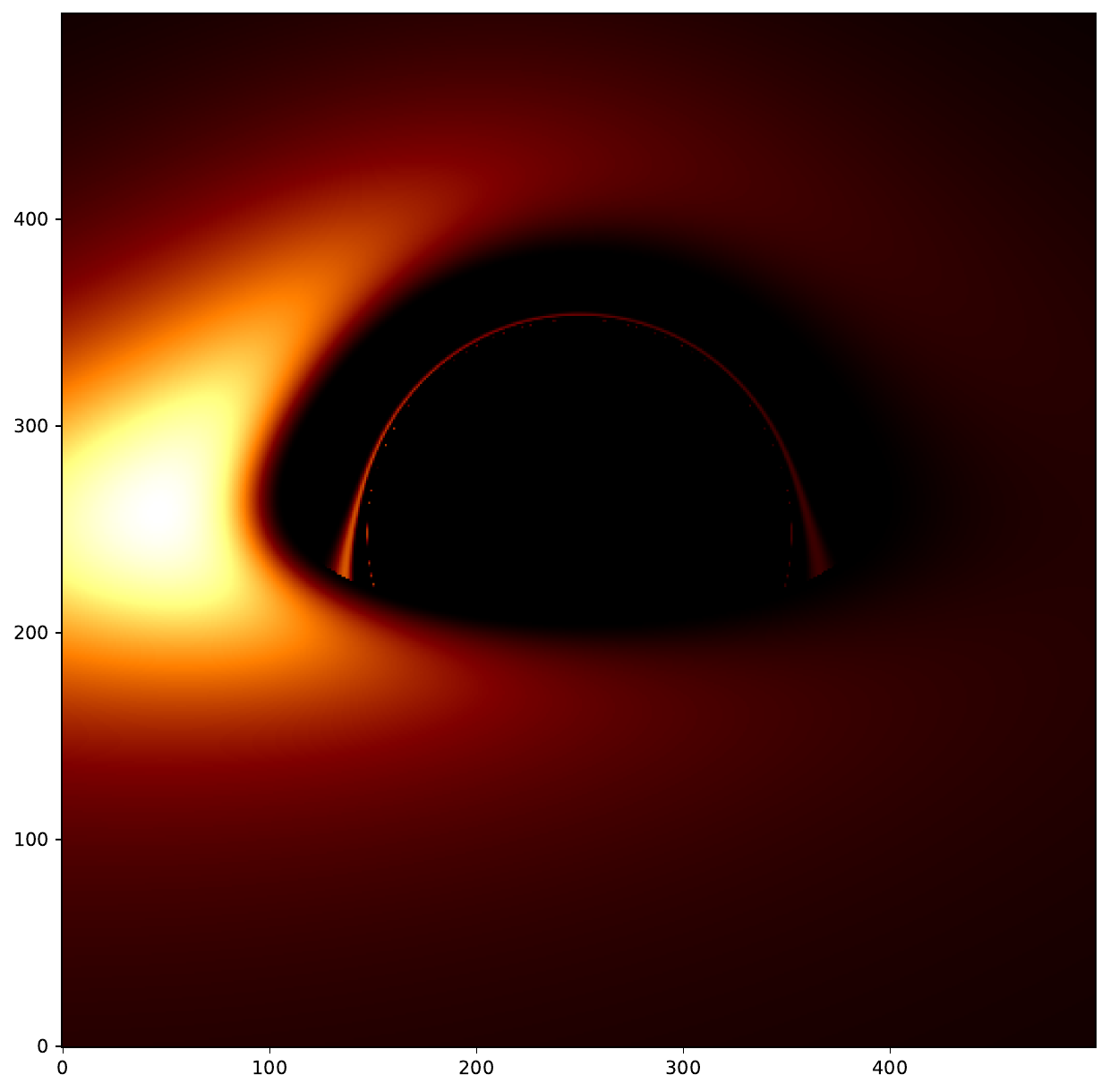} 
        \caption{$\omega=-4/3$}
        \label{14d}
    \end{subfigure}
\hspace{0.001cm} 
    \begin{subfigure}[t]{0.18\textwidth}
        \centering
        \includegraphics[width=\textwidth]{PageThorneGR.pdf} 
        \caption{Schwarzschild}
        \label{14e}
    \end{subfigure}
    \caption{The accretion disks generated using GYOTO code for $\alpha=0.48$.}
    \label{14}
\end{figure}
  \begin{figure}[htbp]
    \centering
    \begin{subfigure}[t]{0.18\textwidth}
        \centering
        \includegraphics[width=\textwidth]{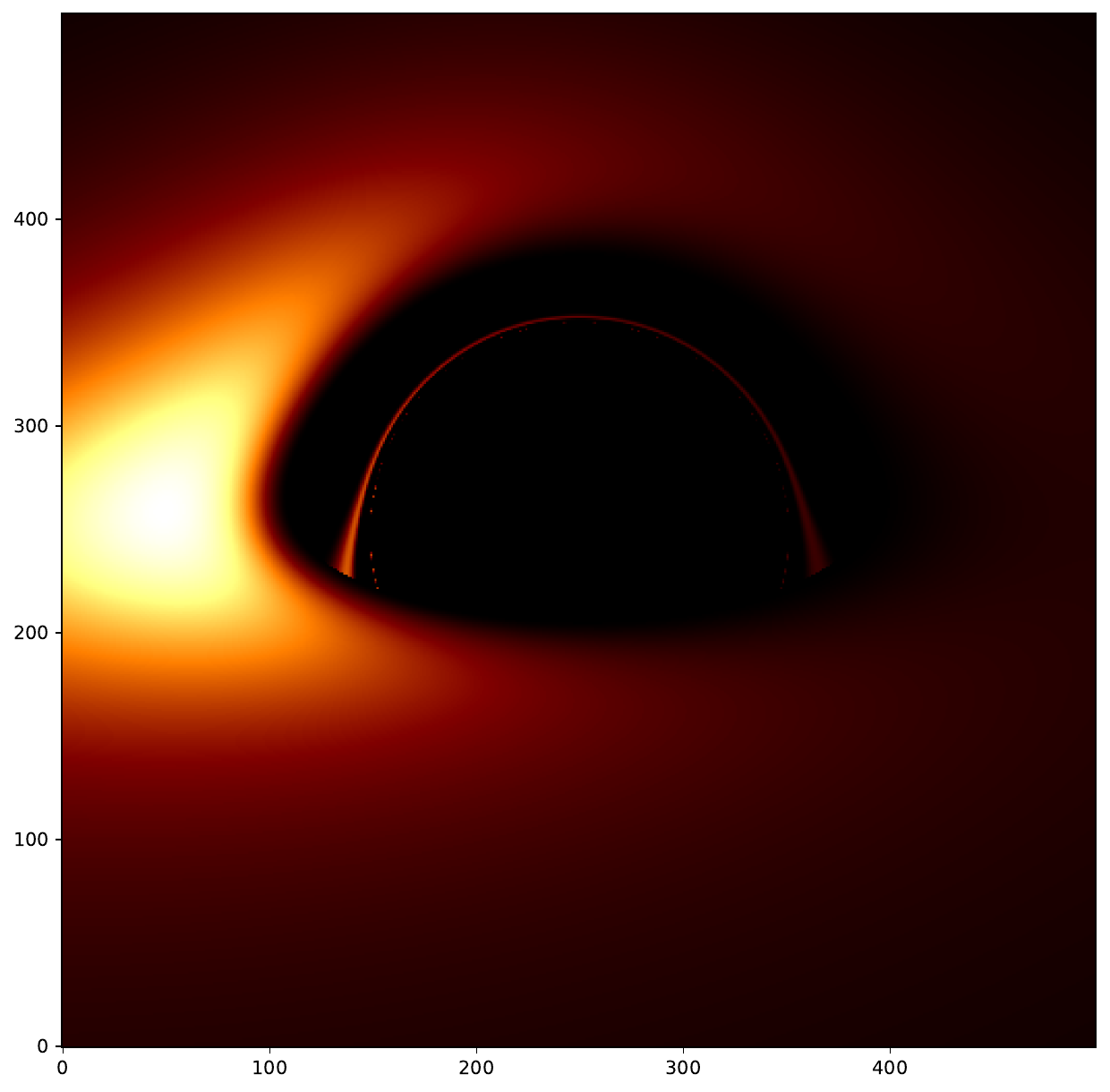} 
        \caption{$\omega=1/3$}
        \label{15a}
    \end{subfigure}
    \hspace{0.001cm} 
    \begin{subfigure}[t]{0.18\textwidth}
        \centering
        \includegraphics[width=\textwidth]{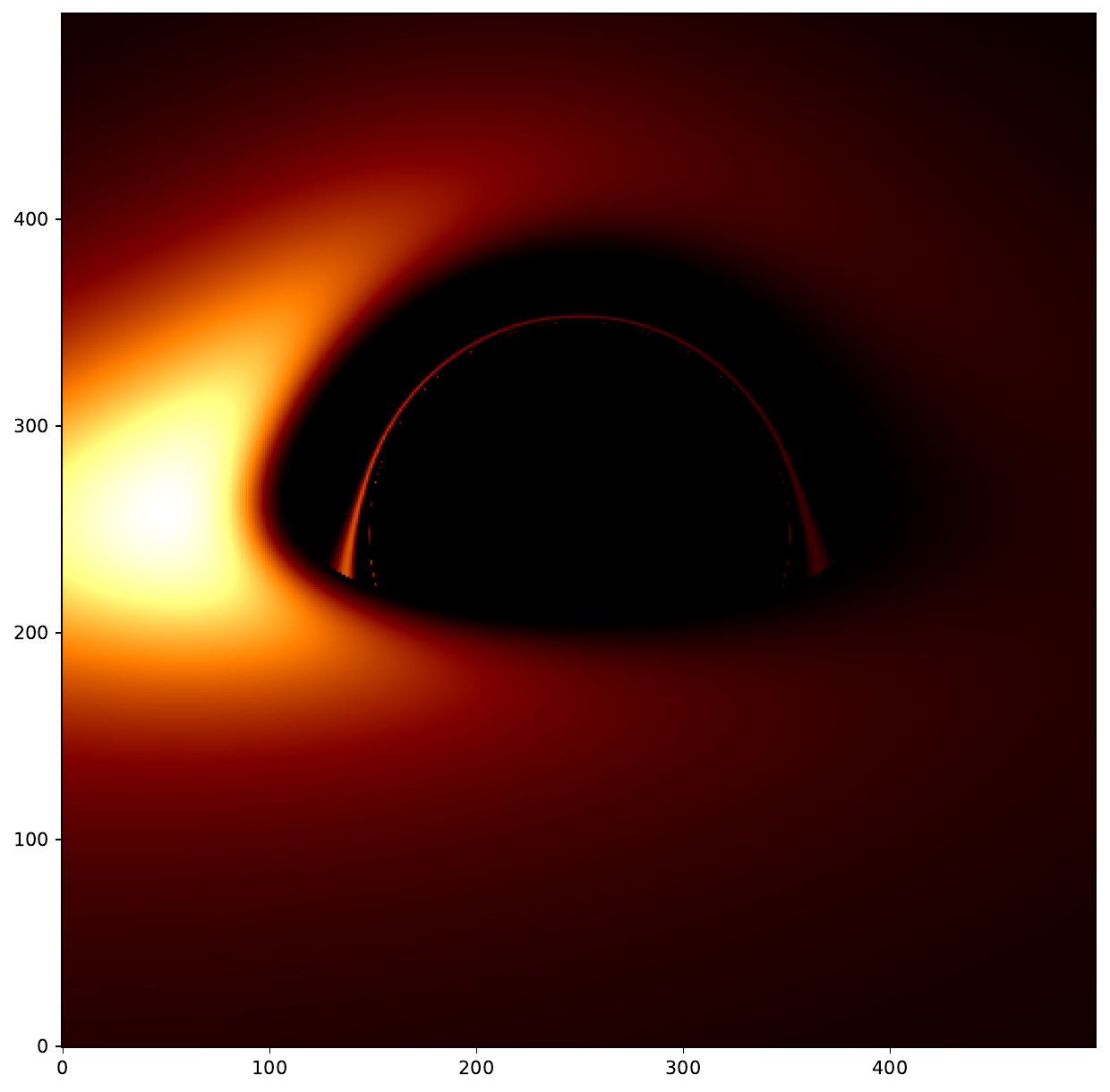} 
        \caption{$\omega=0$}
        \label{15b}
    \end{subfigure}
     \hspace{0.001cm} 
    \begin{subfigure}[t]{0.18\textwidth}
        \centering
        \includegraphics[width=\textwidth]{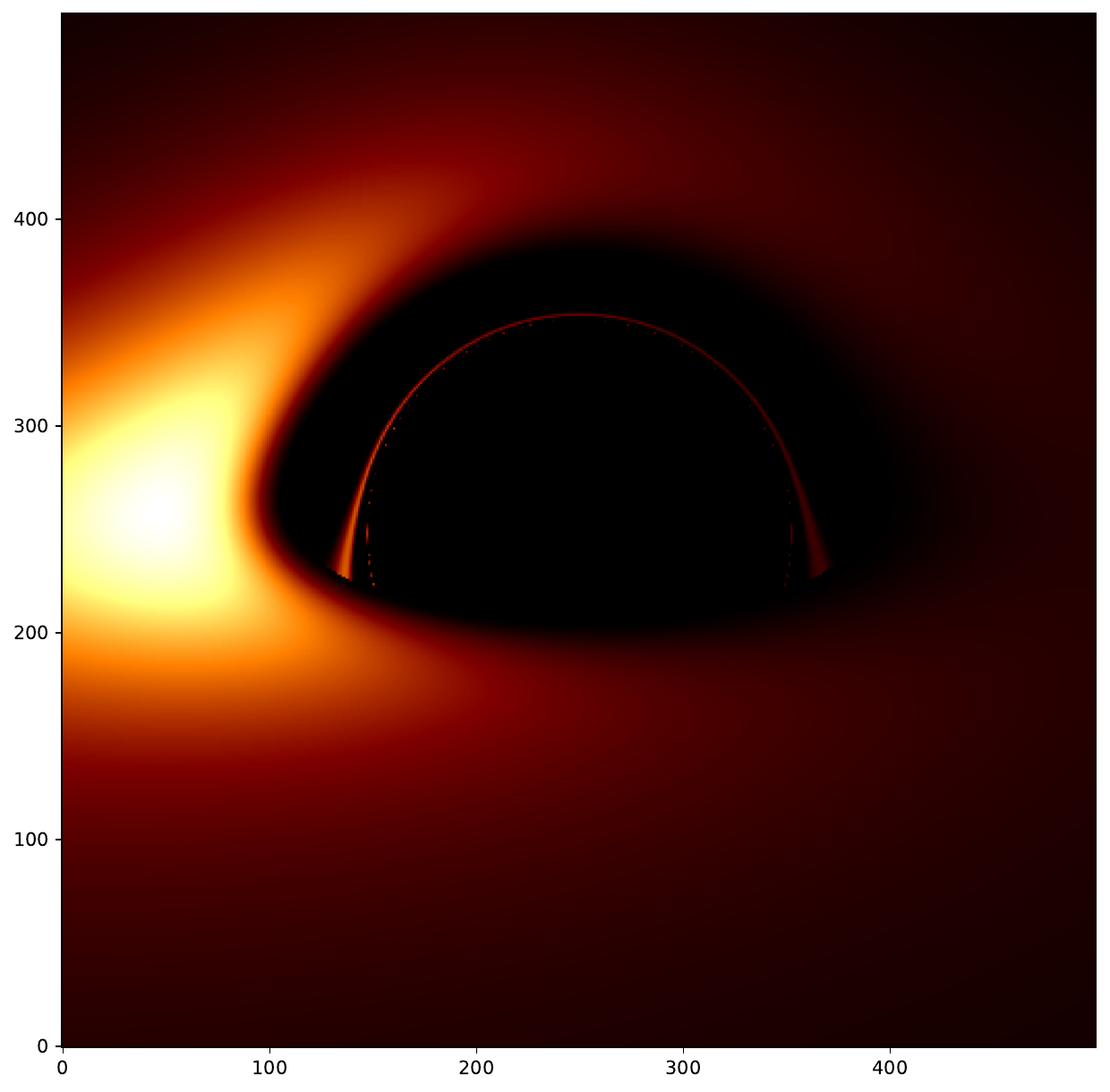} 
        \caption{$\omega=-2/3$}
        \label{15c}
    \end{subfigure}
    \hspace{0.001cm} 
    \begin{subfigure}[t]{0.18\textwidth}
        \centering
        \includegraphics[width=\textwidth]{f1.33330.520-70c.pdf} 
        \caption{$\omega=-4/3$}
        \label{15d}
    \end{subfigure}
\hspace{0.001cm} 
    \begin{subfigure}[t]{0.18\textwidth}
        \centering
        \includegraphics[width=\textwidth]{PageThorneGR.pdf} 
        \caption{Schwarzschild}
        \label{15e}
    \end{subfigure}
    \caption{The accretion disks generated using GYOTO code for $\alpha=0.52$.}
    \label{15}
\end{figure}

The accretion disks in figures \ref{14} and \ref{15} correspond to fixed values of \( \alpha = 0.48 \) and \( \alpha = 0.52 \), respectively. As mentioned earlier, we consider that the metrics are equivalent for different \( \omega \) when \( \alpha = 0.50 \), so the images of the accretion disks for different \( \omega \) are identical in this case. Considering the range of \( \alpha \) for different \( \omega \), we present images of the accretion disk with \( \alpha = 0.48 \) and \( \alpha = 0.52 \). As shown in figure \ref{14}, with \( \alpha = 0.48 \), we observe that the areas of the shadow for \( \omega = 1/3 \), \( \omega = 0 \), \( \omega = -4/3 \), and Schwarzschild are not very different, and are all larger than the area of the shadow with \( \omega = -2/3 \). As shown in figure \ref{15}, with \( \alpha = 0.52 \), we observe that the areas of the shadow for \( \omega = 1/3 \), \( \omega = 0 \), \( \omega = -2/3 \), and Schwarzschild are also not very different, and are all larger than the area of the shadow with \( \omega = -4/3 \).

\section{Conclusions}

We investigated the nature of the thin accretion disk surrounding a QFMGBH by utilizing the Novikov-Thorne model. Based on the two conditions that the QFMGBH spacetime is asymptotically flat and the numerical values of the physical properties of the thin accretion disk are real numbers, we restricted the range of the parameter \( \alpha \) and within which we computed the numerical solution of \( r_{\rm{ISCO}} \). The influence of \( \alpha \) on the energy flux, the radiation temperature, the luminosity spectrum, the energy conversion efficiency of the thin accretion disk, and the shadow of the QFMGBH was discussed  in detail under four specific values of the parameter \( \omega \).

The analysis revealed that the radii of the ISCO for \( \omega = 1/3 \), \( \omega = 0 \), and \( \omega = -2/3 \) increase as \( \alpha \) increases, while the radius of the ISCO for \( \omega = -4/3 \) decreases as \( \alpha \) increases. The findings indicate that as \( \alpha \) increases: for \( \omega = 1/3 \), the energy flux and the radiation temperature decrease, while the luminosity spectrum and energy conversion efficiency first increase slowly and then decrease; for \( \omega = 0 \) and \( \omega = -2/3 \), the energy flux, the radiation temperature, the luminosity spectrum, and the energy conversion efficiency first increase and then decrease; for \( \omega = -4/3 \), the energy flux, the radiation temperature, the luminosity spectrum, and the energy conversion efficiency first increase and then decrease; furthermore, for \( \omega = 1/3 \), \( \omega = 0 \), and \( \omega = -2/3 \), the shadow areas all increase as \( \alpha \) increases, while for \( \omega = -4/3 \), the shadow area decreases as \( \alpha \) increases.

Noticeably, the observable characteristics of the thin accretion disk, including the radius of the ISCO, the maximum values of \( F_\text{max}(r) \), \( T_\text{max}(r) \), and \( \nu L(\nu)_\text{max} \), and the shadow for different \( \omega \), demonstrate that the thin accretion disk around the QFMGBH is smaller than the Schwarzschild BH in GR, but hotter and brighter.  

Taking into account the remarkable differences in the energy flux, the radiation temperature, the luminosity spectrum, the conversion efficiency, and the shadow between the QFMGBH and Schwarzschild BH in GR, these findings suggest the potential for distinguishing between quantum fluctuation modified gravity and standard GR based on observable characteristics.

\begin{acknowledgments}
This study is supported in part by National Natural Science Foundation of China (Grant No. 12333008) and Hebei Provincial Natural Science Foundation of China (Grant No. A2021201034). The GYOTO code can requested from us by email.
\end{acknowledgments}

\bibliographystyle{ieeetr}
\bibliography{ref2}
\end{document}